\documentclass[aps, showpacs, prb,superscriptaddress,twocolumn,a4paper]{revtex4-1}
\usepackage{graphicx}
\usepackage{amsmath, amsfonts, amssymb, bm}
\usepackage{color}
\usepackage{threeparttable}
\usepackage{multirow}
\usepackage[utf8]{inputenc}
\usepackage[english]{babel}
\usepackage{amsmath,array,amssymb,pst-all}
\usepackage{dsfont}
\usepackage{bbold}

\newcommand{\e}{\mbox{e}}

\newcommand{\sugb}[1]{\textcolor{blue}{#1}}

\begin{document}

\title{Photophysical properties of BODIPY-derivatives  for the implementation of organic solar cells: a computational approach}
\author{Duvalier Madrid-Úsuga}
\altaffiliation{{\tt{duvalier.madrid@correounivalle.edu.co}}}
\address{Centre for Bioinformatics and Photonics (CIBioFi), Calle 13 No.~100-00, Edificio E20 No.~1069, Universidad del Valle, 760032 Cali, Colombia}
\address{Quantum Technologies, Information and Complexity Group---QuanTIC,
Departamento de Física, Universidad del Valle, 760032 Cali, Colombia}

\author{Alejandro Ortiz}
\altaffiliation{{\tt{alejandro.ortiz@correounivalle.edu.co}}}
\address{Centre for Bioinformatics and Photonics (CIBioFi), Calle 13 No.~100-00, Edificio E20 No.~1069, Universidad del Valle, 760032 Cali, Colombia}
\address{Heterocyclic Compounds Research Group---GICH, Departamento de Química, Universidad del Valle, 760032, Cali, Colombia}

\author{John H. Reina}
\altaffiliation{{\tt{john.reina@correounivalle.edu.co}}}
\address{Centre for Bioinformatics and Photonics (CIBioFi), Calle 13 No.~100-00, Edificio E20 No.~1069, Universidad del Valle, 760032 Cali, Colombia}
\address{Quantum Technologies, Information and Complexity Group---QuanTIC,
Departamento de Física, Universidad del Valle, 760032 Cali, Colombia}

  
\begin{abstract}
Solar cells based on organic compounds are a proven emergent alternative to conventional electrical energy generation. Here, we provide a computational study of power conversion efficiency optimization of BODIPY-derivatives by means of their associated  open circuit voltage, short-circuit density, and fill factor. In so doing, we compute for the derivatives' geometrical structures, energy levels of frontier molecular orbitals, absorption spectra, light collection efficiencies, and exciton binding energies, via density functional theory (DFT) and time dependent (TD)--DFT calculations. We fully-characterize four D--$\pi$--A  (BODIPY) molecular systems  of high efficiency and improved $J_{sc}$ that are well suited for integration into bulk heterojunction (BHJ) organic solar cells as electron-donor materials in the active layer. Our results are two-fold: We found that molecular complexes  with an structural isoxazoline ring exhibit a higher  power conversion efficiency (PCE), a useful result for improving the BHJ current, and, on the other hand, by considering the molecular systems as electron-acceptor materials, with P3HT as the electron-donor in the active layer, we found a high PCE compound favorability with a pyrrolidine ring in its structure, in contrast to the molecular systems built with an isoxazoline ring. The  theoretical characterization of the  electronic properties of the  BODIPY-derivatives here provided, computed with a combination of ab-initio methods  and quantum models, can be readily applied to other sets of molecular complexes in order to  hierarchize  optimal  power conversion efficiency. 

\end{abstract}
\maketitle

\section{Introduction}

The intensive exploitation of energy resources, the growth of global demand for energy and the associated environmental crisis; e.g., the increase of fossil fuel consumption is a concern that requires immediate attention and evidences the need for alternative sources of renewable energy to avoid further impact on climate change.   In this sense, the metal-free organic solar cells (OSCs) is a potential candidate to aid in this task.  During the past two decades, organic solar cells have attracted much attention due to their  quick improvement of photovoltaic properties and environmental and economic advantages; these include low-cost fabrication, lightweight, mechanical flexibility, high versatility due to applications in many fields, and the simplicity of the  synthesis process~\sugb{\cite{cheng2018, sun2018, li2018}}. The improvement of organic materials  photo-physical capabilities of new  hole transporting material (HTM) and electron transporting material (ETM), which constitute the photo-active components in organic photovoltaic (OPV) cell devices, can increase the performance of such solar energy conversion materials~\sugb{\cite{liao2016, cai2017, cabrera2018,madrid2018,madrid2019}}. Thus, BODIPY-Fullerene derivatives synthesis becomes a valuable alternative to compose hole and  electron transporting materials. This is so, since, as an acceptor component, the Fullerene  promotes the presence of many closely spaced electronic levels and a high degree of delocalized charge within an extended $\pi$-conjugated structure, as well as a low reorganization energy, making these molecular assemblies ideal for acting as electron transporting materials in the charge-separation state~\sugb{\cite{ganesamoorthy2017, sathiyan2016, ware2021}}.
\medskip

Recently, among the commonly used artificial photosynthetic chromophores, the Boron Dipyrromethene (BODIPY) compounds have been of particular interest due to their extraordinary spectral and electronic properties as well as their high absorption coefficients and high emission quantum yields. These translate into an extensive use of BODIPY structures as building blocks for both nano-antenna systems and charge separation units~\sugb{\cite{el2014, guo2014, banuelos2016, bandi2015}}.  The BODIPY dyes open the possibility to extend the absorption cross section over a wide spectral range of OPV devices, e.g., BODIPY-C$_{60}$ dyads~\sugb{\cite{liu2016, gao2017}}.  Consequently, they have been used in applications involving the  development of sensors, photo-dynamic therapy agents, light-energy-harvesting systems~\sugb{\cite{amin2012,shi2013}}, and protein markers, to cite but a few~\sugb{\cite{yee2005,tan2004}}. 
\medskip

In general, the absorption spectra in the near-infrared region motivates the adoption of BODIPY as materials for OSC integration~\sugb{\cite{gautam2017}}. However, few works in the literature show the derivatives of BODIPY and BODIPY-fullerene in the active layer at the construction of photovoltaic devices, among which we find a derivative of BODIPY and PCBM that is used as an active layer in a Heterojunction solar cell, reaching a PCE of 1.34\% in 2009~\sugb{\cite{rousseau2009}}. In 2016, Liao {\it et al.} synthesized and characterized different molecules based on BODIPY and implemented an OSC, using PC$_{61}$BM as electron-acceptor material, with a PCE of 2.15\%~\sugb{\cite{liao2016}}; in 2017, Singh {\it et al.} reported a higher efficiency at 7.2\% for an OSC built with BODIPY-DTF and PC$_{61}$BM~\sugb{\cite{rao2017}}. 
\medskip

Despite the low PCE values for these OSCs compared to those of the amorphous silicon/crystalline silicon heterojunction cells (PCE around 26.63\%), the bulk-heterojunction (BHJ) organic solar cells including conjugated polymers and BODIPY and BODIPY-Fullerene derivatives provide an effective solution for the roll-to-roll production and the fabrication on flexible substrates~\sugb{\cite{heeger2014,huang2014}}.  This said, higher power conversion efficiencies in BHJ solar cells require a morphology that delivers electron and hole percolation pathways to optimize the electronic transport,  including a donor-acceptor contact area high enough to form a charge transfer state near to the unit. This constitutes a significant structural challenge, particularly for semiconductor polymer-fullerene systems~\sugb{\cite{Aren2017}}.  Then, it  becomes a necessity  to provide information on molecular systems within the BHJ organic solar cells which can improve the energy conversion potential in such structures.
\medskip

In this work, we propose novel molecular systems derived  from BODIPY-fullerene (see Figure~\sugb{\ref {Fig_1}}), which exhibit experimental evidence of intramolecular electron transfer processes  as D--$\pi$--A materials~\sugb{\cite{cabrera2017, cabrera2018, calderon2021}}. By means of Scharber's model, we were able to compute the PCE, and to estimate  the maximum conversion efficiency (under ideal conditions),  thus, establishing some working conditions for application to photovoltaic devices.
\medskip

\section{Results and Discussion}

\subsection{Molecular Systems and Computational Results}\label{comp}

\begin{figure}[ht]
\centering
\includegraphics[scale=0.45]{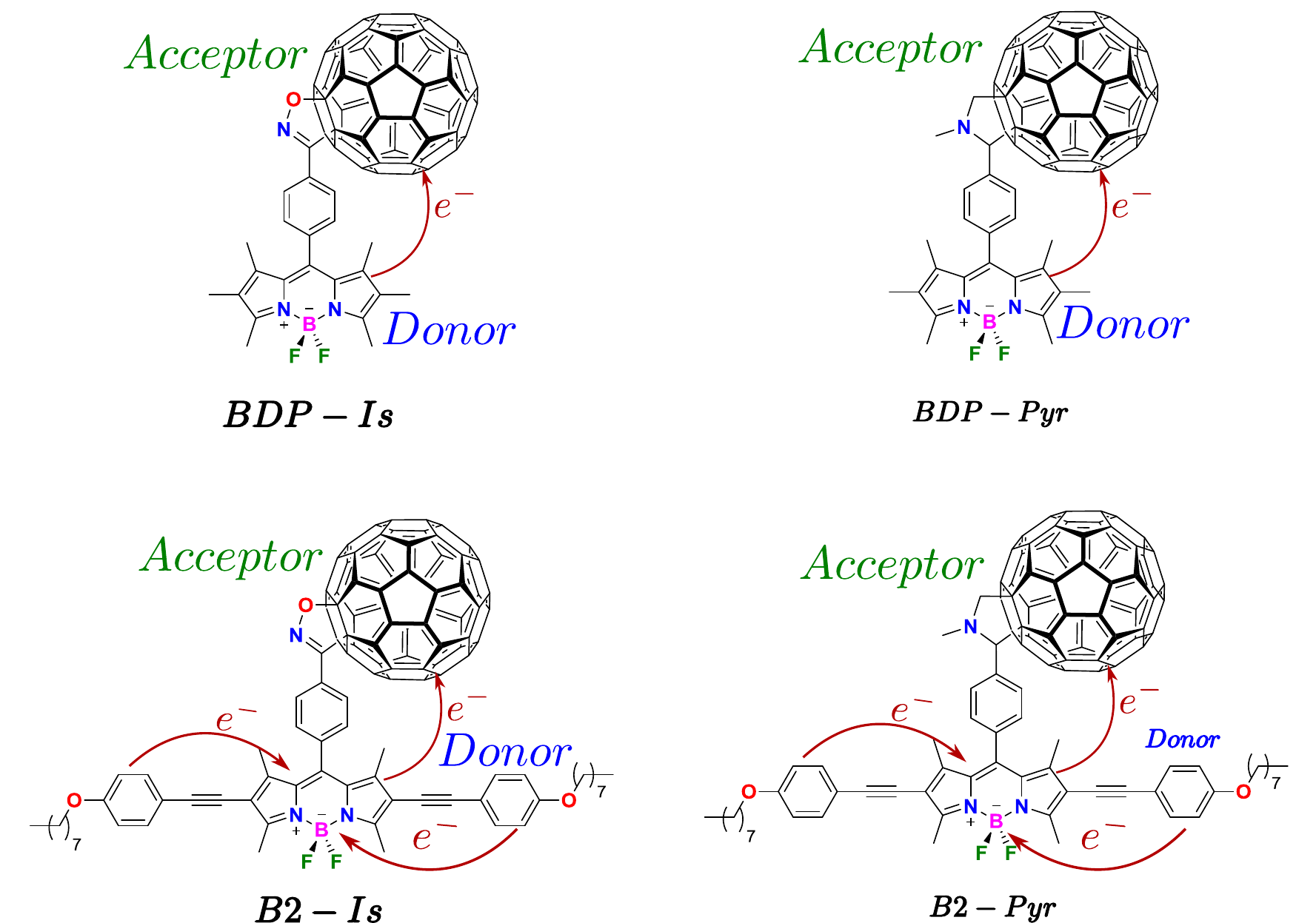} 
\caption{Molecular chromophores under study. For simplicity, we have labeled the complexes as \textbf{\textit{BDP-Is}}, \textbf{\textit{BDP-Pyr}},  \textbf{\textit{B2-Is}} and \textbf{\textit{B2-Pyr}}.}
\label{Fig_1}
\end{figure}

In the development of the criteria here described, we  consider  the molecular systems reported in~\sugb{\cite{cabrera2017,cabrera2018, calderon2021}} (see Figure~\sugb{\ref{Fig_1}}). These exhibit experimental parameters already measured which allow for a direct comparison with our  theoretical results, via the calculation of observables such as electronic energy levels, absorption spectra, charge transfer states, and others.
The molecules under consideration are shown in Figure~\sugb{\ref{Fig_1}}; we have termed them as follows: \textbf{\textit{BDP--Is}}, \textbf{\textit{BDP--Pyr}}, \textbf{\textit{B2--Is}}, and \textbf{\textit{B2--Pyr}}. The calculations involve the use of the density functional theory (DFT) with a  B3LYB exchange-correlation functional~\sugb{\cite{becke1993}} and the base set 6-311G(d,p); these allow  the investigation of the ground state optimization geometry of the electron-acceptor and electron-donor components, and the prediction of the frontier molecular orbital energies~\sugb{\cite{ganji2015,ganji2016}}.
\medskip

The excitation energy, the absorption spectral electronic coupling, the oscillator strength, and the charge transfer state of the systems shown in Figure~\sugb{\ref{Fig_1}}, are based on the  TD-DFT with a hybrid CAM-B3LYP functional. In the ground state, the geometrical configurations were completely optimized with DFT while the excited state was optimized via TD-DFT. The  molecular environment influence was modeled as a dielectric medium by  employing a Conductor-like Polarization Continuum Model (C--PCM) \sugb{\cite{takano2005,chiu2016}}. The C--PCM  is regularly used to recognize the solvation which indicates the solvent effects in the molecular complex.  Here, the polarity effects on the molecular photo-physics are modelled on the basis of considering (polar)  Methanol and (non-polar)  Toluene as dissolvents, which are characterized by the dielectric constants $\varepsilon_M=32.6130$ and $\varepsilon_T=2.3741$, respectively.
\medskip

There are different methods to calculate the electronic properties in organic molecules. Here, we specialize to the DFT Kohn--Sham energy levels, while the HOMO level can be related to the ionization potential according to Janak's theorem~\sugb{\cite{janak1978}}. Nevertheless, we remark that the results will markedly rely on the employed functional.

\subsection{Transport Properties} \label{Charge_transfer}

\textbf{\textit{Charge Transfer States.}} The charge transfer (CT) states, between the electron--donor and electron--acceptor parts in the molecular system, have a fundamental role in organic solar cells operation. Thus, a better understanding of CT  intramolecular processes is expected to optimize the organic photovoltaic (OPV) materials, improving the device performance \sugb{\cite{vandewal2016, su2016}} towards the CT states control. Although electronic excitations on individual molecules are well characterized, a thorough  description of CT states is still missing. The challenge to analyze the CT states arises from the multiple factors dependence, such as the molecular geometry, the nature of the pure and mixed donor/acceptor domains, and the interaction with the surrounding polar or non-polar environment. At the electronic structure level, all these factors can influence the electron polarization and extend the electronic delocalization having a strong influence on the energy and nature of the CT state \sugb{\cite{zheng2017}}.
\medskip

The intramolecular electron transfer process involves the following: the system is initially photo-excited from the ground electronic state to $\pi$--$\pi^{*}$ states ($\vert \psi_D^j \rangle$) primarily localized on either place in the donor part. A  non-radiative CT process then occurs at a later stage, corresponding to an electronic transition from the $\vert \psi_D^j \rangle$ state to a CT state ($\vert \psi_A^i \rangle$)  involving a  significant CT from donor to acceptor. 
Once such electron dynamics  is known, we employ the density functional theory (DFT) to determine the electronic ground state, and the time-dependent density functional theory (TD--DFT) to calculate the excited states as well as the geometry and energy by the electrostatic environment affected in CT states. From a quantum mechanically viewpoint, the understanding of this kind of process may help at improving the power conversion efficiency. Our principal concern is then to observe and understand the effects due to the solvent polarity and geometry of the molecular complexes on the electronic and transport properties, especially in excitation and CT state, which currently is an open problem.
\medskip

\begin{table}[ht]
\begin{center}
\centering
\caption{Most probable donor ($D$) or acceptor ($A$) fragments  for localization of electronic orbitals in the molecular systems \textbf{\textit{BDP-Is}}, \textbf{\textit{BDP-Pyr}}, \textbf{\textit{B2-Is}}, and \textbf{\textit{B2-Pyr}}.}
\scalebox{1.0}{
\begin{tabular}{ccccccccccc} 
\hline
\hline
\\
\multicolumn{1}{m{2cm}}{} & \textbf{\textit{BDP-Is}} & \textbf{\textit{BDP-Pyr}} & \textbf{\textit{B2-Is}} & \textbf{\textit{B2-Pyr}}
\\ 
\hline 
\hline
\\
\multicolumn{1}{m{2cm}}{\centering \bf Orbitals} &  &  & \bf Fragment
\\
\hline
\hline
 HOMO-5  & A & A & A & A \\ 
 HOMO-4  & A & A & A & A \\  
 HOMO-3  & A & A & A & A \\
 HOMO-2  & A & A & A & A \\
 HOMO-1  & D & D & D & D \\
 HOMO    & D & D & D & D \\
 LUMO    & A & A & A & A \\
 LUMO+1  & A & A & A & A \\
 LUMO+2  & A & A & A & D \\
 LUMO+3  & D & D & D & A \\
 LUMO+4  & A & A & A & A \\
 LUMO+5  & A & A & A & A \\
\hline
\end{tabular}}
\label{Tab_1}
\end{center}  
\end{table}

\noindent
As an overview of the molecules with the performed calculations, Table~\sugb{\ref{Tab_1}} shows the frontier orbital localization according to the donor and acceptor fragments  for the molecular complexes \textbf{\textit{BDP-Is}}, \textbf{\textit{BDP-Pyr}}, \textbf{\textit{B2-Is}}, \textbf{\textit{B2-Pyr}}. All the molecular systems have a similar distribution of orbital localizations, as shown in Table~\sugb{\ref{Tab_1}}.  The molecular orbitals from \textit{HOMO--2} to \textit{HOMO--5} are located in the electron-acceptor part while the \textit{HOMO--1} and HOMO are in the electron-donor part, and the bound \textit{LUMOs} are in the electron-acceptor part.  Nevertheless, the \textit{LUMO+3} in the \textbf{\textit{BDP-Is}}, \textbf{\textit{BDP-Pyr}}, \textbf{\textit{B2-Is}} and \textit{LUMO+2} in the \textbf{\textit{B2-Pyr}} system are in the confined electron-donor fragment.  With the orbital localizations, we observe that a low-energy transition leading by the formation of the  $\vert\psi_D^j\rangle$ state located in the electron-donor  takes place from \textit{HOMO} or \textit{HOMO--1} to \textit{LUMO+3} (or \textit{LUMO+2} in the \textbf{\textit{B2-Pyr}} molecule); the  $D^*$--$\pi$--$A$ represents the resulting state. On the other hand, an electronic transition in a highly probable CT takes place from one of the bound \textit{HOMOs} located at the electron-donor to the \textit{LUMO} located at the electron-acceptor giving rise to the  $\vert\psi_A^i\rangle$ state as $ D^{+}$--$\pi$--$A^-$ represented.
\medskip

\begin{figure*}[ht]
\centering
\includegraphics[scale=0.25]{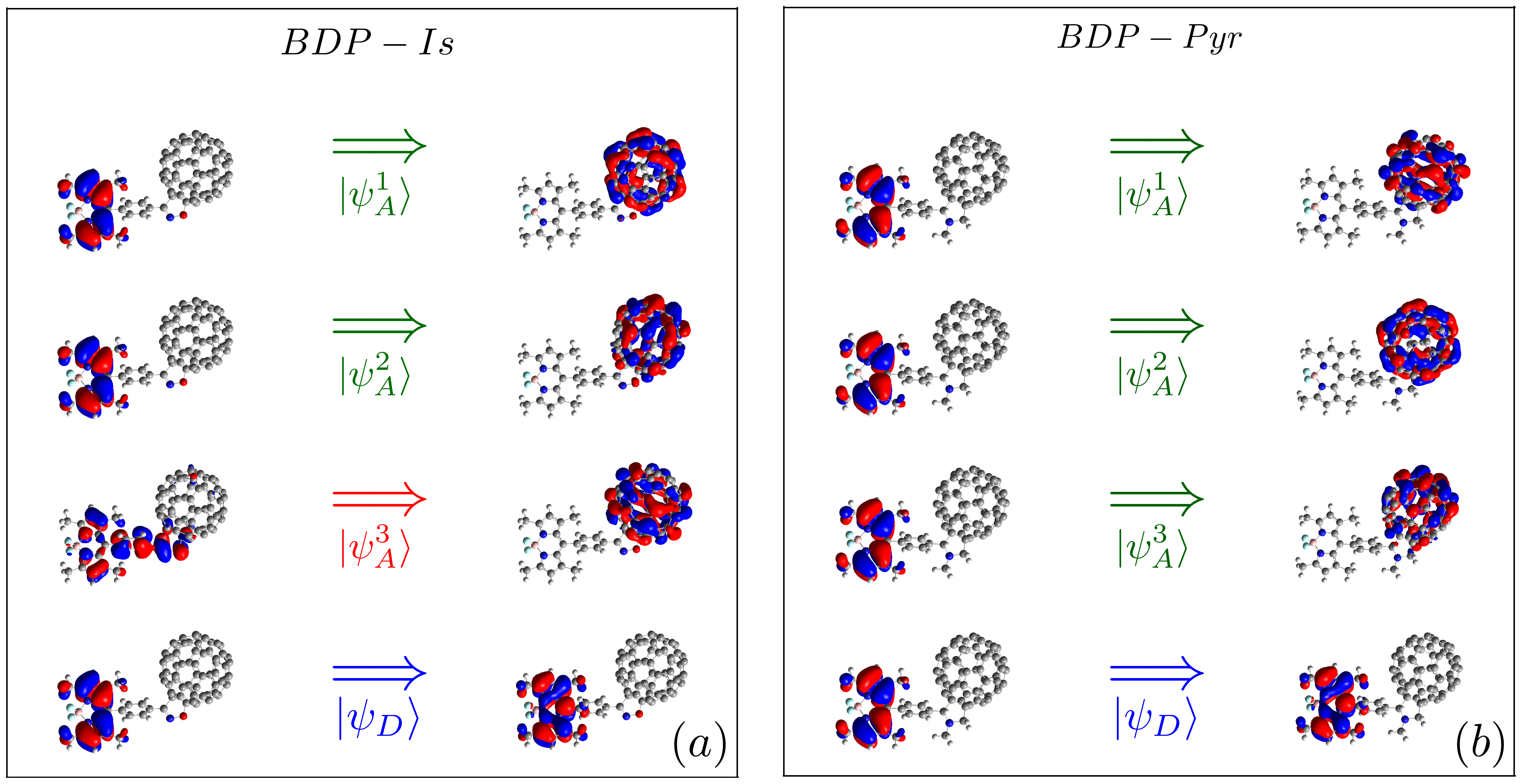}
\caption{Electron density associated with the excited states for the molecular complexes (a) \textbf{\textit{BDP-Is}} and (b) \textbf{\textit{BDP-Pyr}}  in Methanol solvent.}
\label{Fig_2}
\end{figure*}

\noindent
Table~\sugb{S1} summarizes the first twenty five lowest-energy electronic transitions with their energies, oscillator strength, and orbital involved in the electronic transition for the molecular systems \textbf{\textit{BDP-Is}}, \textbf{\textit{BDP-Pyr}}, \textbf{\textit{B2-Is}}, and \textbf{\textit{B2-Pyr}} in Methanol (for Toluene, see Table~\sugb{S2} in the  \sugb{\textit{Supplementary Information}}).  We consider an electronic state with excitation energies below 3.7~eV, to match the spectral range of OPV devices. We characterize the different states by considering their detachment-attachment electron densities that are illustrated in Figure~\sugb{\ref{Fig_2}} (for the \textbf{\textit{B2-Pyr}} and \textbf{\textit{B2-Is}} see Figure~\sugb{S1} in \sugb{\textit{Supplementary Information}}), following the procedure by Head-Gordon \sugb{\cite{head1995}}.
\medskip

\noindent
In the \textbf{\textit{BDP-Is}} and \textbf{\textit{BDP-Pyr}} configurations, we find four relevant excited states, which can be classified as follows:

\begin{itemize}
\item  The excited states  $\vert\psi_D^j\rangle$ are localized on the donor part, and present an  oscillator strength with small charge transfer values between donor and acceptor fragments, and can be  identified as a donor $\pi$--$\pi^*$ excitation.

\item  The excited states labelled as $\vert\psi_A^1\rangle$, $\vert\psi_A^2\rangle$, ...,$\vert\psi_A^i\rangle$  are characterized by a  small oscillator strength, and a large charge transfer values between the  donor and the  acceptor part.  This process is essentially due to the transference of one electron from the donor to the acceptor segment, constituting a charge transfer state.
\end{itemize}
\noindent

As for the charge transfer state energies used in the charge transfer process, they are calculated with a combination of excitation energies from TD-DFT calculations, implementing the ionization potential ($IP$) and electronic affinity ($EA$) of the different charge sites and a Coulomb energy term that describes the interaction of the relevant charge sites. The ground state is the reference state and its diabatic state energy is set to zero. The energy of the locally excited diabatic state ($D^*$--$A$) is calculated as the vertical excitation energy to the adiabatic state that most represents it. The energy of one of the charge transfer states (CTS) with a positive charge at site $x$ and a negative charge at site $y$ is calculated as~\sugb{\cite{blomberg1998, powell2017}}:
\begin{equation}
E_{exc_j}=IP_D-E_{\hbar\nu}-EA_j+E_c(r_{xy}),
\end{equation}
where $IP_D$ is the ionization potential of the donor fragment, which was calculated by subtracting the total energy of the neutral molecule from that of the cation, as calculated by DFT. $EA_j$ is the electronic affinity of the acceptor fragment  and was obtained by subtracting the total energy of the anion from that of the neutral state. $E_{\hbar\nu}$ is the lowest single excited state energy of the donor molecule calculated by TD-DFT, and $E_c(r_{xy})$ is the Coulomb interaction between the cation on the donor fragment  and the anion on the acceptor fragment. $r_{Dj}$ is the distance from the center of mass of the donor fragment  to the center of mass of the acceptor fragment,  based on the optimized ground state structure.
\medskip

\begin{figure*}[ht]
\centering
\includegraphics[scale=0.9]{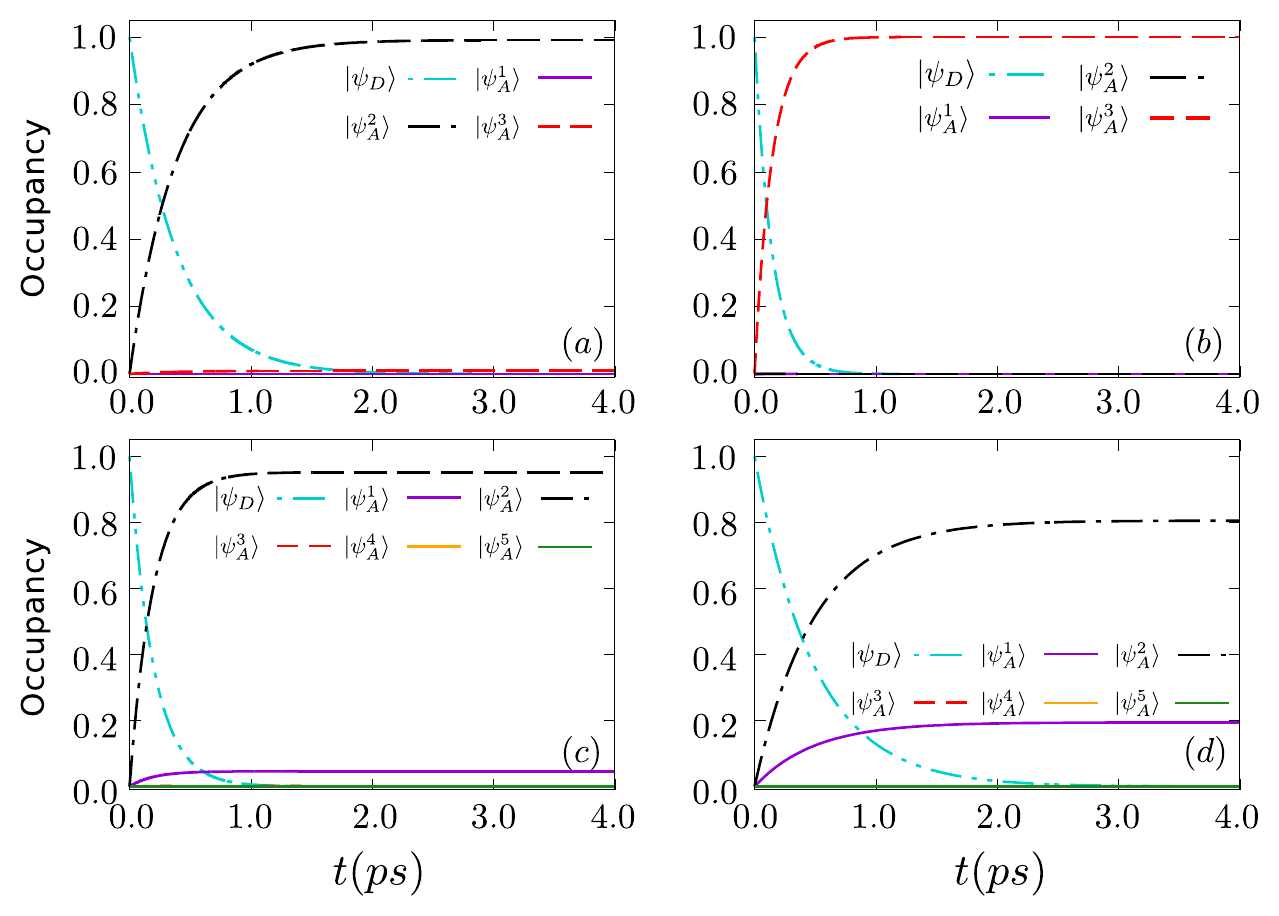}
\caption{Occupancy of individual excited states as a function of time for  (a)  \textbf{\textit{BDP-Is}}, (b)  \textbf{\textit{BDP-Pyr}}, (c)  \textbf{\textit{B2-Is}}, and (d)  \textbf{\textit{B2-Pyr}}, in Methanol.}
\label{Fig_3}
\end{figure*}

Therefore, we found one excitation state $\vert\psi_D^1\rangle=\vert\psi_D\rangle$ for the complexes \textbf{\textit{BDP-Is}} and \textbf{\textit{BDP-Pyr}}.  With $f_{osc}=0.6040$ for the ~\textbf{\textit{BDP-Is}} system and $f_{osc}=0.6068$ for the~\textbf{\textit{BDP-Pyr}} complex, with resulting excitation energy of 2.4811~eV and 2.5042~eV, respectively, mainly originated from a \textit{HOMO}--to--\textit{LUMO+3} transition, corresponding to the  $\vert\psi_D\rangle$ ($\pi$--$\pi^{*}$) state in the BODIPY molecules. In addition, there are three CT states characterized by a small oscillator strength (0.0001$-$0.0092), as shown in Figure~\sugb{\ref{Fig_2}}. In the molecules \textbf{\textit{B2-Is}} and \textbf{\textit{B2-Pyr}}, we found one excited state $\vert\psi_D\rangle$ with $f_{osc}=0.7820$ (system~\textbf{\textit{B2-Is}}) and $f_{osc}=0.7800$ (system~\textbf{\textit{B2-Pyr}}), and five CT states $\vert\psi_A^i\rangle$ with small oscillator strengths (0.0001$-$0.0025) and large electron transfer (0.95\textit{e}$-$1.2\textit{e}) (see Figure \sugb{\ref{Fig_2}}). However, the charge transfer states used in this work are those with a favorable driving force ($\Delta G<0$), since they present a greater interaction with the excited states and higher electron transfer as shown in Figure~\sugb{\ref{Fig_3}}.
\medskip

According to the above, in the Tables~\sugb{\ref{Tab_3}} and Table~\sugb{S1} (\sugb{\textit{Supplementary Information}}), the system~\textbf{\textit{BDP-Is}} has two CT states, $D^+$--$A^-$. The first one with excitation energy 1.9715~eV, $\vert\psi_A^1\rangle$, is formed through the first electronic transition due to the main contribution of \textit{HOMO}-to-\textit{LUMO+1}. The second one, $\vert\psi_A^2\rangle$, is formed via the \textit{HOMO}-to-\textit{LUMO+2} electronic transition characterized by an excitation energy of 2.1632~eV and also a vanishing oscillator strength. Thus, there are two channels for the dissociation of the exciton $D^*$--$A$ into $D^+$--$A^-$ via the photo-induced states $\vert\psi_A^1\rangle$ and $\vert\psi_A^2\rangle$ with a favorable driving force $\Delta G_{\vert\psi_A^1\rangle}=-0.5096$~eV and $\Delta G_{\vert\psi_A^2\rangle}=-0.3179$~eV, respectively. The third state $\vert\psi_A^3\rangle$ is not considered here because of a  disfavorable driving force $\Delta G=0.0851$~eV. According to Veldman {\it et al.}, $\Delta G_{\vert \psi_A^i \rangle}$ is the dissipated energy in the CT state formation ($\vert \psi_D\rangle \rightarrow \vert \psi_A^i\rangle $); a negative value for $\Delta G_{\vert \psi_A^i \rangle}$ ensures the necessary driving force for the photoinduced electron transfer (PET) in a photovoltaic blend~\sugb{\cite{veldman2009}}. Considering the first twenty lowest energy values for electronic transitions, we found only two CT states with a favorable driving force in the molecular structures \textbf{\textit{B2-Is}}  and \textbf{\textit{B2-Pyr}}.  However, in all the systems, we found $\vert\psi_A^i\rangle$ states with disfavored driving forces ($\Delta G>0$), as shown  in Table~\sugb{\ref{Tab_3}}; this can be attributed to the lack of interactions between frontier molecular orbitals.
\medskip

Comparing the energy amount of the $\big\vert\psi_{D(A)}^{j(i)}\big\rangle$ states between molecules  with similar geometry (\textbf{\textit{BDP-Is}} and \textbf{\textit{BDP-Pyr}}), we found a considerable difference probably attributed to the conjugation of electrons in the ion pair of the oxygen ($O$) and nitrogen ($N$) atoms within the isoxazoline fragment with the $\pi$-conjugated BODIPY core system.
\medskip

The addition of the alkoxyphenylethynyl fragments in the molecules \textbf{\textit{B2-Is}} and \textbf{\textit{B2-Pyr}} generate excited states with lower energy than those presented in \textbf{\textit{BDP-Is}} and \textbf{\textit{BDP-Pyr}} systems.  The energetic ordering of solvated states in Methanol is maintained for all systems, and the favorable driving forces follow the hierarchy  $E_D>E_A^1>E_A^2>E_A^3$ because the solvation energy increases with the strength of the dipole, as expected.
\medskip

On the other hand, the electronic coupling forces calculated between the excitation state $\vert\psi_D^j\rangle$ and the state $\vert\psi_A^i\rangle$, obtained via the Generalized Mulliken-Hush (GMH), as shown in Table \sugb{\ref{Tab_3}}, suggest as a possible mechanism the charge transfer from the excited state $\vert\psi_D\rangle$ to the $\vert\psi_A^i\rangle$ states of photo-induced charge. The electronic coupling between the $\vert\psi_D\rangle$ and $\vert\psi_A^2\rangle$ states is greater for the \textbf{\textit{BDP-Pyr}} molecule than for the other systems. The coupling between the $\vert\psi_D\rangle$ and $\vert\psi_A^2\rangle$ maintained this behavior, as can be seen in Table~\sugb{\ref{Tab_3}} for the electron transfer case.

\begin{table}[htp] 
\begin{center}
\centering
\caption{Electronic coupling ($V_{ji}$), reorganization energy ($\lambda_e$), driving force ($\Delta G$), activation energy ($E_r$), and  constant rates ($s^{-1}$) according to  Marcus theory for ($j\rightarrow i$) electronic transitions for the molecular complexes \textbf{\textit{BDP-Is}}, \textbf{\textit{BDP-Pyr}}, \textbf{\textit{B2-Is}}, and \textbf{\textit{B2-Pyr}}, in Methanol.}
\scalebox{0.720}{
\begin{tabular}{ccccccc} 
\hline
\hline
\\
\multicolumn{1}{m{1.8cm}}{\centering \textbf{Molecule}} & $i\rightarrow j$ & $\vert V_{ji}\vert$ ($m$eV) & $\lambda_{e}$ (eV) & $\Delta G$ (eV) & $E_r$ (eV) & $\kappa_e(s^{-1})$
\\
\hline 
\hline
\\
\textbf{\textit{BDP-Is}}  & $\vert\psi_D\rangle\rightarrow\vert\psi_A^1\rangle$   &  7.36 & 0.1347 & -0.5096 & 0.2609 & $1.0212\times 10^{08}$ \\ 
                          & $\vert\psi_D\rangle \rightarrow \vert\psi_A^2\rangle$ & 25.17 & 0.1347 & -0.3179 & 0.0623 & $2.5945\times 10^{12}$ \\  
                          & $\vert\psi_D\rangle \rightarrow \vert\psi_A^3\rangle$ &  4.17 & 0.1347 &  0.0851 & 0.0897 & $2.4735\times 10^{10}$ \\
                          &                                                       &                &                  &                        \\  
\textbf{\textit{BDP-Pyr}} & $\vert\psi_D\rangle \rightarrow \vert\psi_A^1\rangle$ & 29.33 & 0.1356 & -0.5245 & 0.2788 & $8.0932\times 10^{08}$ \\
                          & $\vert\psi_D\rangle \rightarrow \vert\psi_A^2\rangle$ &  6.22 & 0.1356 & -0.4399 & 0.1707 & $2.3846\times 10^{09}$ \\
                          & $\vert\psi_D\rangle \rightarrow \vert\psi_A^3\rangle$ & 19.01 & 0.1356 & -0.2448 & 0.0220 & $7.0189\times 10^{12}$ \\
                          &                                                       &                &                  &                        \\  
\textbf{\textit{B2-Is}}   & $\vert\psi_D\rangle \rightarrow \vert\psi_A^1\rangle$ &  4.87 & 0.1332 & -0.2785 & 0.0396 & $2.3546\times 10^{11}$ \\ 
                          & $\vert\psi_D\rangle \rightarrow \vert\psi_A^2\rangle$ & 12.04 & 0.1332 & -0.1988 & 0.0081 & $4.8650\times 10^{12}$ \\
                          & $\vert\psi_D\rangle \rightarrow \vert\psi_A^3\rangle$ &  1.41 & 0.1332 &  0.0436 & 0.0586 & $9.3726\times 10^{09}$ \\
                          & $\vert\psi_D\rangle \rightarrow \vert\psi_A^4\rangle$ &  1.16 & 0.1332 &  0.1036 & 0.1052 & $1,0618\times 10^{09}$ \\
                          & $\vert\psi_D\rangle \rightarrow \vert\psi_A^5\rangle$ &  2.87 & 0.1332 &  0.1837 & 0.1884 & $2.5744\times 10^{08}$ \\
                          &                                                       &                &                  &                        \\                   
\textbf{\textit{B2-Pyr}}  & $\vert\psi_D\rangle \rightarrow \vert\psi_A^1\rangle$ &  3.70 & 0.1358 & -0.2159 & 0.0118 & $3.9816\times 10^{11}$ \\
                          & $\vert\psi_D\rangle \rightarrow \vert\psi_A^2\rangle$ &  6.20 & 0.1358 & -0.1108 & 0.0011 & $1.6507\times 10^{12}$ \\
                          & $\vert\psi_D\rangle \rightarrow \vert\psi_A^3\rangle$ &  1.50 & 0.1358 &  0.1366 & 0.1366 & $5.3790\times 10^{08}$ \\
                          & $\vert\psi_D\rangle \rightarrow \vert\psi_A^4\rangle$ &  1.90 & 0.1358 &  0.1723 & 0.1748 & $1.9255\times 10^{08}$ \\
                          & $\vert\psi_D\rangle \rightarrow \vert\psi_A^5\rangle$ &  3.00 & 0.1358 &  0.2763 & 0.3127 & $2.2841\times 10^{06}$ \\
\hline
\end{tabular}}
\label{Tab_3}
\end{center}
\end{table}
\medskip

\textbf{\textit{Charge Transfer Rate Constants.}} The kinetics of the photoinduced charge transfer is modeled following the image of Marcus's theory \sugb{\cite{marcus1993, marcus1964}}. The charge transfer is a crucial process involved in many physical and biological phenomena (e.g., in a first approximation,  the photosynthesis) \sugb{\cite{gray2009}}. Marcus' result gives:
\begin{equation}
\textcolor{black}{k_e=\frac{2\pi}{\hbar}\frac{\vert V_{ij}\vert^{2}}{\sqrt{4\pi\lambda_e k_BT}}\exp{\bigg[-\frac{(\Delta G+\lambda_e)^{2}}{4\lambda_ek_{B} T}\bigg]}},
\label{Ecu4_1C}
\end{equation}
where $V_{ij}$ is the electronic coupling between the states $\vert\psi_D^j\rangle$ and  $\vert\psi_A^i\rangle$, $\lambda_e$ is the reorganization energy and $k_{B}$ is the Boltzmann constant. 

We consider a scenario where the photoexcitation towards the excited state (i.e., large oscillator strength states) is followed by nonradiative transitions. Those transitions are to lower-lying charge transfer state corresponding to charge separation instantaneously upon absorption (e.g., states $\vert\psi_A^1\rangle$, $\vert\psi_A^2\rangle$ and$\vert\psi_A^3\rangle$ in the \textbf{\textit{BDP-Is}} configuration). Marcus' constant rate  for the $\vert\psi_D^j\rangle$ to $\vert\psi_A^i\rangle$ state transition and the parameters that influence these constant rates  (the electronic coupling coefficients $V_{ij}$, the reorganization energy $\lambda_e$, the driving force $\Delta G=E_A^i-E_D^j$, and the activation energies $E_r=\frac{(\Delta G+\lambda_e)^2}{4\lambda_e}$) are given in Table \sugb{\ref{Tab_3}}.
\medskip

For the molecular complexes, the electron transfer rates $\vert\psi_D\rangle\rightarrow\vert\psi_A^2\rangle$ have the same order of magnitude in all systems, with the exception of gthe  \textbf{\textit{BDP-Pyr}} system where it is presented in $\vert\psi_D\rangle\rightarrow\vert\psi_A^3\rangle$. Additionally,  we found that the CT state with a value of $\Delta G>0$ for the photoinduced electron transfer process has an ET lower rate  (between two to six magnitude orders) than the CT states with $\Delta G<0$ (a  favorable driving force), giving us information about the relevant CT states for  the charge transfer analysis.
\medskip

We noted that $\lambda_e<\vert\Delta G\vert$ for the transitions $\vert\psi_D\rangle\rightarrow \vert\psi_A^1\rangle$, and $\vert\psi_D\rangle \rightarrow \vert\psi_A^2\rangle$ in the molecular systems \textbf{\textit{BDP-Is}}, \textbf{\textit{BDP-Pyr}} and \textbf{\textit{B2-Is}}. However, this condition is only accomplished by the $\vert\psi_D\rangle \rightarrow \vert\psi_A^1\rangle$ state in the \textbf{\textit{B2-Pyr}} system, implying that the  $\vert\psi_A^2\rangle$ for \textbf{\textit{B2-Pyr}} takes place in the inverted region of Marcus~\sugb{\cite{yin2012}}. Additionally, the $\vert\psi_D\rangle \rightarrow \vert\psi_A^2\rangle$ transition in the \textbf{\textit{B2-Pyr}} system shows a $\lambda_e$ slightly larger than $\Delta E$, resulting in an electronic transition with a low activation energy.  Under these conditions, the classic barrier crossing may become more favorable than nuclear tunnelling as a dominant transition mechanism.
\medskip

\textbf{\textit{Charge Transfer Kinetics.}} The overall CT kinetics involves a transition between states $\vert\psi_D^j\rangle$  and $\vert\psi_A^i\rangle$.  Assuming that each of these transitions can be by defined as a Marcus constant rate, as above, one can establish the overall kinetics in terms of a master equation describing the incoherent motion through time-dependent occupation probabilities, $P_i(t)$, of some quantum states, $\vert \psi_A^i \rangle$. Then, the $P_i(t)$ are a solution of rate equations of the type:
\begin{equation}
\dot{P}_i(t)=\sum_{j\neq i}\big[-k_{j\rightarrow i}P_i(t)+k_{i\rightarrow j}P_j(t)\big].
\label{Ecu4_2C}
\end{equation}
This equation contains the rates (of probability transfer per unit time) $k_{ij}$ for transitions from $\vert\psi_D \rangle$ to $\vert \psi_A^i \rangle$. The first term of the right-hand side describes the decrease of $P_i$ in time due to probability transfer from $\vert \psi_D \rangle$ to all other states, and the second term accounts for the reverse process, including the transfer from all other states $\vert \psi_A^i \rangle$ to the  $\vert \psi_D \rangle$ state. In 1928,  Eq.~\sugb{\ref{Ecu4_2C}} was ``intuitively derived'' by W. Pauli~\sugb{\cite{pauli1928}}, then, this expression is frequently called the Pauli Master Equation or just the Master Equation. By considering $k_{i\rightarrow j}=0$, we have
\begin{equation}
P_{\psi_A^i}=\frac{k_{\psi_D\rightarrow \psi_A^i}P_{\psi_D}(0)}{K}\big( 1-\e^{-Kt}\big).
\label{Ecu4_6C}
\end{equation}

Here, $P_{\psi_{A(D)}^i}(t)$ is the excited state population $\vert\psi_{A(D)}^i\rangle$ at time $t$, such that $ P_{\psi_D}(t)+\sum_{i}P_{\psi_A^i}(t)=1$, with $k_e=k_{\psi_D\rightarrow\psi_A^i}$ as the Marcus constant rate  for the electronic transition from $\vert\psi_D^j\rangle$ state  to $\vert\psi_A^i\rangle$  state (see Table~\sugb{\ref{Tab_3}}). We do not include transitions from the excited to the ground state in the master equation, and assume the initial electronic state corresponding to photo-excitation at time $t=0$. Then the occupancy is given by:

\begin{equation}
P_{\psi_A^i}(t=0)=\frac{\vert f_{osc_{\psi_A^i}} \vert}{\sum_i\vert f_{osc_{\psi_A^i}}\vert},
\label{Ecu4_7C}
\end{equation}
where $f_{osc_{\psi_A^i}}$ is the oscillator strength of excited state $\vert\psi_A^i\rangle$. Namely, we impose direct photo-excitation of the interfacial dimer which guarantees that the initial state is  dominated by bright states.
\medskip

\medskip

Figure~\sugb{\ref{Fig_3}} shows the occupancy of each excited state individually as a function of time in Methanol for all the molecular configurations. The initial states are the excited $\pi$--$\pi^*$ states ($\vert\psi_D\rangle$), and the transitions to other states $\vert\psi_A^i\rangle$ occur selectively from the $\vert\psi_D\rangle$ in different time scales. For the \textbf{\textit{BDP-Is}} molecular system, Figure~\sugb{\ref{Fig_3} (a)}, the two CT states ($\vert\psi_A^1\rangle$ and $\vert\psi_A^2\rangle$) begin with zero population and  $\vert\psi_A^1\rangle$ reach their steady-state after $\sim 2.50$~ps. The occupancy steady-state value of $\vert\psi_A^2\rangle$ is larger than $\vert\psi_A^1\rangle$ state, as there are more pathways ($\pi$-delocalization link) leading to $\vert\psi_A^2\rangle$ state than to  $\vert\psi_A^1\rangle$ (see Table~\sugb{\ref{Tab_3}}). However, for the molecular system \textbf{\textit{B2-Is}} we find that the steady state is reached in a lower timescale ($\sim 0.80$~ps), comparing with the ~\textbf{\textit{BDP-Is}} system mostly due to the addition of fragments of the alkoxyphenylethynyl group.
\begin{figure*}[ht]
\centering
\includegraphics[scale=0.9]{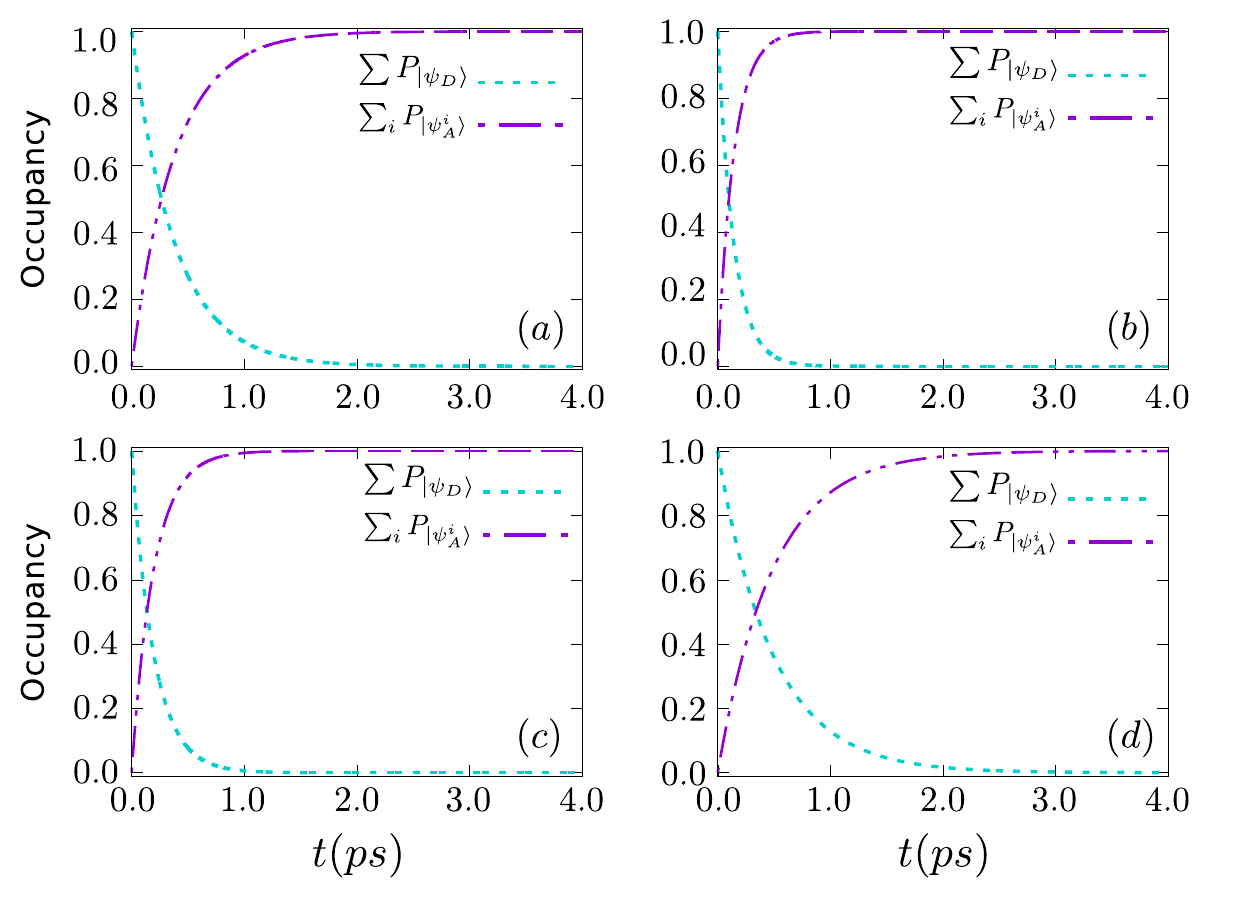}
\caption{Total occupancy for $\vert \psi_D \rangle$ state and the charge transfer state ($\vert \psi_A^i\rangle$) as a function of time for the different molecular complexes: (a)  \textbf{\textit{BDP-Is}}, (b)  \textbf{\textit{BDP-Pyr}}, (c)  \textbf{\textit{B2-Is}} and (d)  \textbf{\textit{B2-Pyr}}, in Methanol.}
\label{Fig_4}
\end{figure*}
The Figure~\sugb{\ref{Fig_3} (c)} and~\sugb{(d)} shows occupancy of the individual excited state as a function of time for ~\textbf{\textit{B2-Is}} and~\textbf{\textit{B2-Pyr}} molecules. For example, the $\pi$--$\pi^*$ state ($\vert\psi_D\rangle$) is strongly coupled to the $\vert\psi_A^2\rangle$ state while is weakly coupled to the $\vert\psi_A^1\rangle$ state, in the molecular system~\textbf{\textit{B2-Is}}. As a result, the $\vert\psi_A^2\rangle$ state occupancy quickly rises after photoexcitation remaining, essentially, constant after that at the branching ratio value, $P_{\psi_D}(0)k_{\psi_D\rightarrow \psi_A^2}/K$. However, for the~\textbf{\textit{B2-Pyr}} molecular system, the $\vert\psi_D\rangle$ state is mostly stronger coupled to the $\vert\psi_A^2\rangle$ state than to the $\vert\psi_A^1\rangle$ state. The latter should not be confused with the occupancy equilibrium of the $\vert\psi_A^i\rangle$ state, which will be obtained in a slower time scale only. The total occupancy of the $\vert\psi_D\rangle$ states and of the $\vert\psi_A^i\rangle$ states are plotted in Figure~\sugb{\ref{Fig_4}}. The overall time scale on which the occupancy of excited state depleted and that on which the occupancy of CT states increases in the \textbf{\textit{BDP-Is}} configuration is very similar to that observed in the \textbf{\textit{BDP-Pyr}} configuration (a similar process occurs between \textbf{\textit{B2-Is}} and \textbf{\textit{B2-Pyr}}). The CT occurs on the subpicosecond time scale, with 90$\%$ of the charge transferred in $\sim 0.80$~ps for the  \textbf{\textit{BDP-Is}},  $\sim 0.40$~ps for the~\textbf{\textit{BDP-Pyr}} system and $\sim 0.50$~ps for the  \textbf{\textit{B2-Is}} system. A higher transfer time was found for the  \textbf{\textit{B2-Pyr}} system observed with a  95$\%$ of the transferred charge  in $\sim 1.8$~ps.
\medskip

Figure~\sugb{\ref{Fig_3}} shows that CT states with driving forces $\Delta G>0$ are unfavorable for the CT process, and have a very small (or null) occupation when considering the dynamics of the transfer rate. However, the states with $\Delta G<0$ present a dynamic of considerable importance, since the processes of charge transfer present between the involved states with this condition are mainly favored. Moreover, these last states have the highest transfer rate as can be seen in the Table~\sugb{\ref{Tab_3}}.
\medskip

\textbf{\textit{Solvent Polarity Effects.}} The research of bathochromic changes in the electronic spectra of molecules provides information about molecule-solvent interactions. The observed bathochromic shift due to the increasing solvent polarity  depends on the difference between the permanent dipole moments of the ground and excited state, in agreement with the dielectric polarization theory. This theory states that the bigger the moment difference between dipoles of the ground state and excited state, the  higher the spectral shift by the solvent-induced~\sugb{\cite{lee1988}}. When the dipole moment of the excited state is larger than that of the ground state, solute-solvent interactions of the excited state are stronger than those of the ground state and a red-shift of maximum absorption  will be observed.
\medskip

\begin{figure*}[ht]
\centering
\includegraphics[scale=0.9]{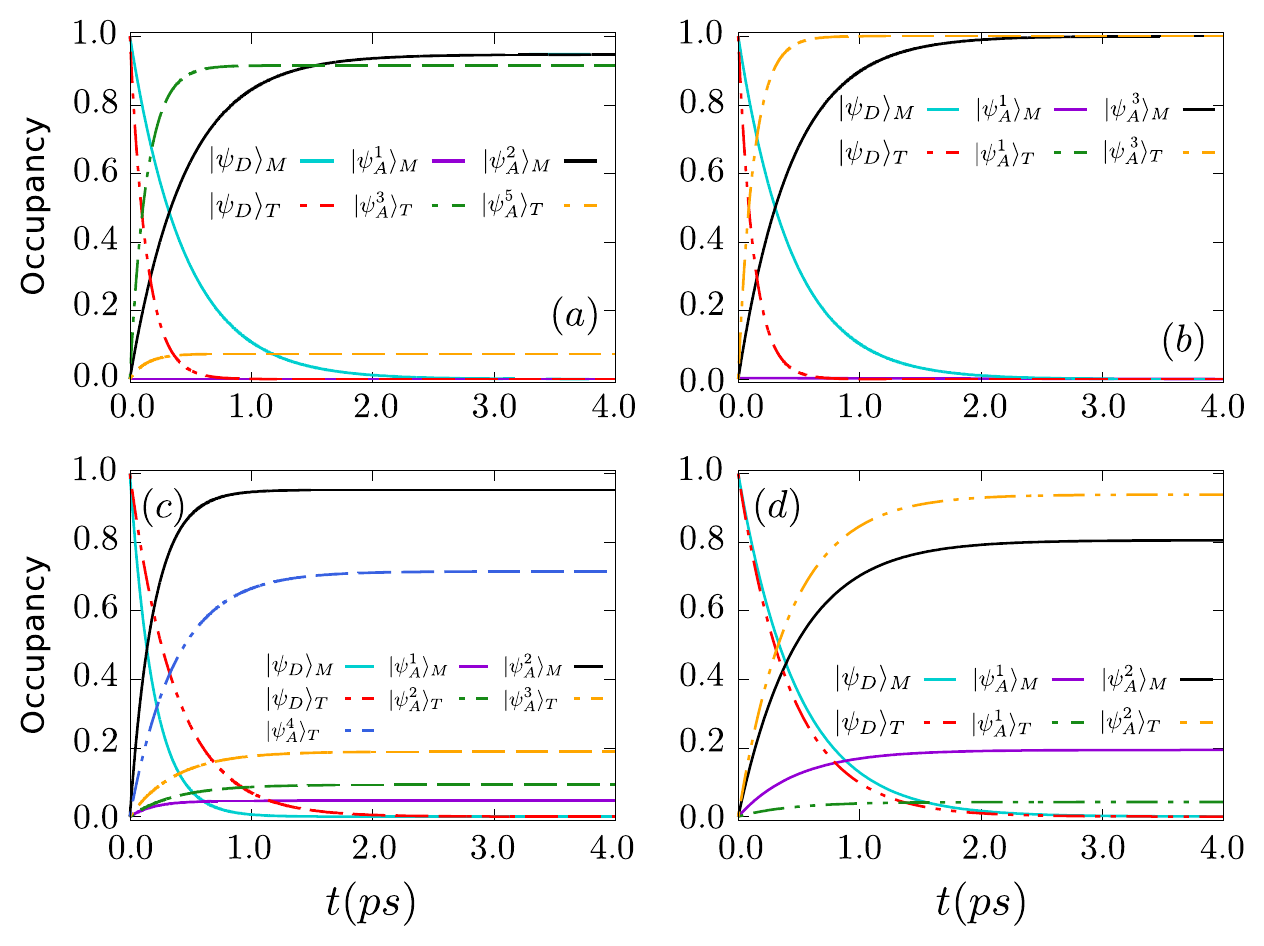}
\caption{Occupancy of individual $\pi$--$\pi^*$ ($\vert \psi_D\rangle$) and $\vert \psi_A^i\rangle$ states for (a) \textbf{\textit{BDP-Is}}, (b)~\textbf{\textit{BDP-Pyr}}, (c)~\textbf{\textit{B2-Is}}, and (d)~\textbf{\textit{B2-Pyr}} configurations in  Methanol ($M$) and Toluene ($T$). The dashed lines are the $\vert \psi_D\rangle_T$ and $\vert \psi_A^i\rangle_T$ states for Toluene, and the solid line is the $\vert \psi_D\rangle_M$ and $\vert \psi_A^i\rangle_M$ for Methanol.}
\label{Fig_5}
\end{figure*}

\begin{figure*}[ht]
\centering
\includegraphics[scale=0.9]{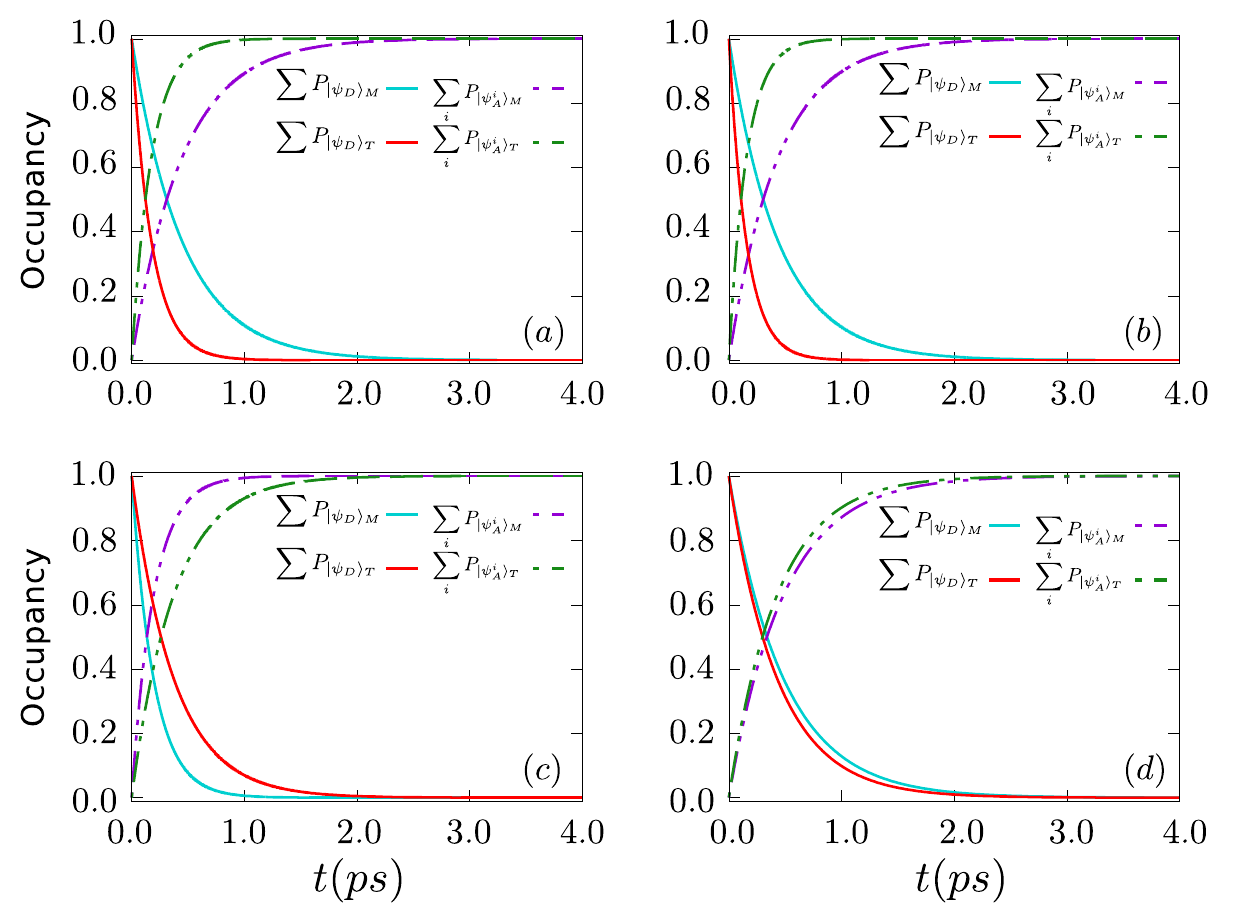}
\caption{Total occupancy dynamics for the $\vert \psi_D \rangle$ state and the charge transfer state ($\vert \psi_A^i\rangle$) as a function of time (in Methanol and Toluene) for: (a)~\textbf{\textit{BDP-Is}}, (b)~\textbf{\textit{BDP-Pyr}}, (c)~\textbf{\textit{B2-Is}}, and (d)~\textbf{\textit{B2-Pyr}}.}
\label{Fig_6}
\end{figure*}

To understand the effects of the solvent polarity on the excitation and charge transfer energy values, we studied the bathochromic change of the compounds in Methanol and Toluene as  solvents. Initially, by considering the Toluene as solvent (non-polar), it is observed the additional generation of two charge transfer states with a favored driving force for the~\textbf{\textit{B2-Is}} system regarding the case where Methanol is considered, with energies $E_A^1=1.5979$~eV, $E_A^2=1.6794$~eV, $E_A^3=2.9223$~eV, and $E_A^4=1.9980$~eV. One additional charge transfer state is observed for the~\textbf{\textit{B2-Pyr}} complex in Toluene compared with Methanol with $E_A^1=1.7106$~eV, $E_A^2=1.8143$~eV, and $E_A^3=2.0657$~eV (\sugb{\textit{see Supplementary Information}}). However, for  the \textbf{\textit{BDP-Is}}, and ~\textbf{\textit{BDP-Pyr}} systems the same amount of CT state is maintained with energies $E_A^1=1.7486$~eV,  $E_A^2=1.8169$~eV, and $E_A^3=2.0756$~eV for the \textbf{\textit{BDP-Is}} system and $ E_A^1=1.8685$~eV, $E_A^2=1.9508$~eV, and $E_A^3=2.1516$~eV for the \textbf{\textit{BDP-Pyr}} system. Indeed, there is a decrease in the energy for the excited state $\pi$--$\pi^*$ ($\vert \psi_D \rangle$) in each of the studied systems with values of 2.4811~eV, 2.5042~eV, 2.0868~eV, and 2.1011~eV for the \textbf{\textit{BDP-Is}}, \textbf{\textit{BDP-Pyr}}, \textbf{\textit{B2-Is}}, and \textbf{\textit{B2-Pyr}} molecular systems, respectively.
\begin{table}[htp]
\begin{center}
\centering
\caption{Dipole moments calculated for ground and excited states of \textbf{\textit{BDP-Is}}, \textbf{\textit{BDP-Pyr}}, \textbf{\textit{B2-Is}}, and \textbf{\textit{B2-Pyr}} molecular systems in Toluene and Methanol solvents (in Debye units).}
\scalebox{0.95}{
\begin{tabular}{ccc|cccccccc} 
\hline
\hline
\\
\multicolumn{1}{m{1.5cm}}{\centering Molecule} & \multicolumn{1}{m{1.5cm}}{\centering Ground}  & \multicolumn{1}{m{1.5cm}}{\centering Excited} & \multicolumn{1}{m{1.5cm}}{\centering Ground} & \multicolumn{1}{m{1.5cm}}{\centering Excited} 
\\
\hline
\hline
\\
& \multicolumn{2}{m{2cm}}{\centering Toluene} & \multicolumn{2}{m{2cm}}{\centering Methanol} & 
\\
\hline
\hline
 \textbf{\textit{BDP-Is}}   & 3.8368 & 4.7372  & 3.1336 &  5.3112 \\ 
 \textbf{\textit{BDP-Pyr}}  & 5.6067 & 6.0955  & 4.5418 &  7.4045 \\  
 \textbf{\textit{B2-Is}}    & 3.8467 & 4.6521  & 3.7648 &  5.3845 \\
 \textbf{\textit{B2-Pyr}}   & 4.2827 & 5.3459  & 4.2287 &  6.5372 \\
\hline
\hline
\end{tabular}}
\label{Tab_4}
\end{center}  
\end{table}
\noindent
On the other hand, the dipolar moments of the excited state for the molecular systems interacting with both Methanol and Toluene have larger values than those for the ground state, as seen in Table~\sugb{\ref{Tab_4}}. Hence, the interaction between the molecular complexes and the solvents are greater in the excited state than in the ground state. Complementary, we observed the dipole moments for  Methanol with larger values than those for Toluene in the excited state, indicating a `better' interaction between the molecules and the methanol, therefore, a better stabilization of the \textit{HOMO-LUMO} orbitals in Methanol. From Figure~\sugb{\ref{Fig_5}} and Figure~\sugb{\ref{Fig_6}}, it is clear that the charge transfer process will perform better in the presence of the polar solvent Methanol than in Toluene.
\medskip
The reorganization energy estimation for the \textbf{\textit{BDP-Is}} molecular complex  yields  $\lambda=0.1347$~eV in Methanol and 0.1564~eV in Toluene. The difference between the two can be traced back to the larger amount and extent of charge transfer in Methanol compared to Toluene.
\medskip
\medskip
The existence of intramolecular charge transfer mechanisms in the current molecular systems gives us information about the capabilities to use them as electron-donor materials. However, the analysis of other properties such as the optical, electronic and photovoltaic properties employing the Scharber's model offers us a precise scenario about the potential application of these systems in the construction of  organic solar cells.  Scharber's model describes an estimation for the power conversion efficiency, in bulk heterojunction solar cells, knowing the energy levels (to estimate the capability of the molecule in a BHJ OSC), and provides an indication on the capabilities of the molecular system that is to be achieved by assuming an efficient charge, absorption, and charge separation process.

\subsection{Optical and Electronic Properties} \label{Optical}
\begin{figure}[ht]
\begin{center}
\includegraphics[scale=0.68]{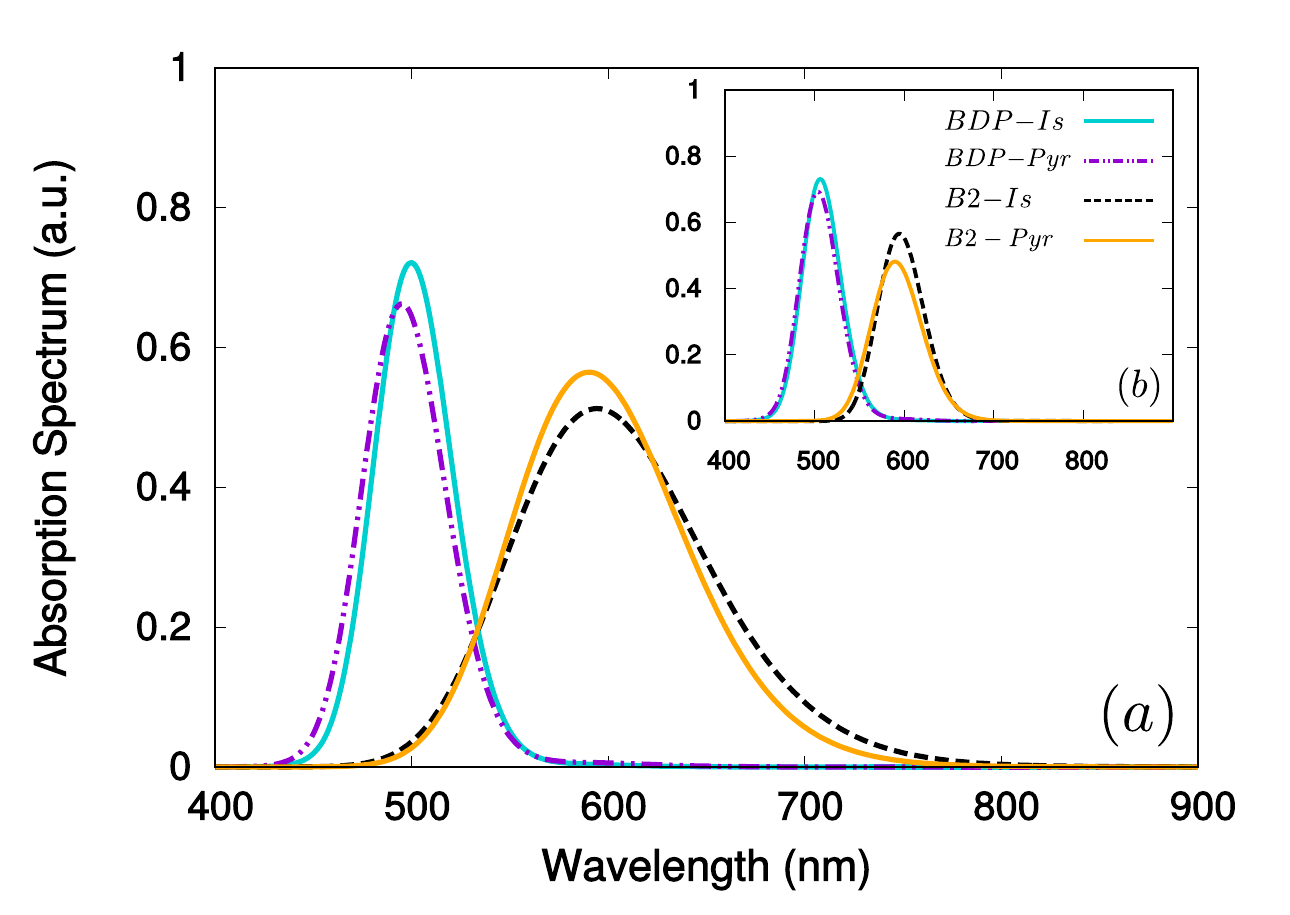}
\caption{Simulated absorption spectra in: (a) Toluene (C$_7$H$_8$) and (b) Methanol (CH$_3$OH) with TD-DFT/CAM-B3LYP/6-311G(d,p) for \textbf{\textit{BDP-Is}}, \textbf{\textit{BDP-Pyr}}, \textbf{\textit{B2-Is}}, and \textbf{\textit{B2-Pyr}} molecular complexes.}
\label{Fig4_8C}
\end{center}
\end{figure}

In OSCs photo-active materials, the active layer (electron-donor material) plays an essential role in the sunlight absorption because the electron-acceptor (e.g., PCBM) should have a weak absorption in the visible and near-infrared regions. To investigate the absorption properties of these molecules, such as excitation energies, oscillators strength of the electronic excitations, the composition of vertical transitions and the UV/VIS absorption spectrum, we carried out TD-DFT calculations with a CAM-B3LYP functional.
\begin{table}[htp] 
\begin{center}
\centering
\caption{Summary of the maximum theoretical   $\lambda_{max}$ and experimental $\lambda_{max}^{^{Exp}}$ absorption wavelength, the excitation energy $E_{exc}$, the oscillator strength $f_{osc}$, the contribution of the most probable transition, and the light harvesting efficiency ($LHE$) of the studied compounds.}
\scalebox{0.78}{
\begin{tabular}{ccccccccc} 
\hline
\hline
\\
\textbf{Molecule} & \textbf{$\lambda_{max}$} (nm) & \textbf{$\lambda_{max}^{^{Exp}}$} (nm) & \textbf{\textit{E}$_{exc}$} (eV) & \textbf{\textit{f}$_{osc}$} & \textbf{Composition} & \textbf{\textit{LHE}} \\
\\ 
\hline
\hline
\large{\text{Toluene}}  &         &  &       &       &                            &       \\
\textbf{\textit{BDP-Is}}  & 508.196 & 506 & 2.440 & 0.654 & H$\rightarrow$L+3   (98\%) & 0.778 \\
\textbf{\textit{BDP-Pyr}} & 503.564 & 505 & 2.462 & 0.652 & H$\rightarrow$L+3   (97\%) & 0.776 \\ 
\textbf{\textit{B2-Is}}   & 599.410 & 591 & 2.068 & 0.821 & H$\rightarrow$L+3   (97\%) & 0.856 \\ 
\textbf{\textit{B2-Pyr}}  & 596.508 & 589 & 2.078 & 0.848 & H$\rightarrow$L+3   (98\%) & 0.853 \\
                          &         &     &       &       &                            &       \\
\hline
\hline
\large{\text{Methanol}} &         &    &       &       &                            &       \\
\textbf{\textit{BDP-Is}}  & 499.775 & -- & 2.481 & 0.602 & H$\rightarrow$L+3   (97\%) & 0.751 \\
\textbf{\textit{BDP-Pyr}} & 495.159 & -- & 2.505 & 0.606 & H$\rightarrow$L+3   (98\%) & 0.753 \\ 
\textbf{\textit{B2-Is}}   & 594.200 & -- & 2.087 & 0.782 & H$\rightarrow$L+3   (97\%) & 0.824 \\ 
\textbf{\textit{B2-Pyr}}  & 590.170 & -- & 2.101 & 0.780 & H$\rightarrow$L+2   (98\%) & 0.823 \\
\hline
\hline
\end{tabular}}
\label{Tab4_5C}
\end{center}
\end{table}
\medskip
\noindent
The excitation energy and  corresponding oscillator strength associated to $\lambda_{max}$ for each molecular system are listed in Table \sugb{\ref{Tab4_5C}}. The spectroscopic parameters corresponding to the D--$\pi$--A derivatives are summarized in Table~\sugb{\ref{Tab4_5C}}, and the simulated absorption curves  for Toluene and Methanol are presented in Figure~\sugb{\ref{Fig4_8C}}. The time-dependent (TD) DFT calculation for the molecular complexes shows that for the \textbf{\textit{BDP-Is}} compound, the optically allowed electronic transition is related to populating the \textit{HOMO}$\rightarrow$\textit{LUMO+3} excitation with high oscillator strength ($f_{osc}$), which is related to an energy band that registers an absorption peak at $508.196$~nm for Toluene and $499.775$~nm for Methanol in the absorption spectrum of  Figure~\sugb{\ref{Fig4_8C}}; possibly ascribed to the intramolecular charge transfer in the BODIPY part. The \textbf{\textit{BDP-Pyr}} compound shows an absorption peak at $503.564$~nm for Toluene, and $495.159$~nm for Methanol associated to the \textit{HOMO}$\rightarrow$\textit{LUMO+3} electronic transition in accordance with the experimentally reported value~\sugb{\cite{cabrera2017}} ($505$~nm for Toluene) and an oscillator strength $f_{osc}=0.6061$. The simulated absorption spectra show, for the ~\textbf{\textit{B2-Is}} compound, a peak with maximum absorption at $599.410$~nm, with $f_{osc}=0.8214$ for Toluene, and $594.200$~nm for Methanol in which the electronic transitions \textit{HOMO}$\rightarrow$\textit{LUMO+3} (representing a $97$\%) correspond to an intramolecular charge transfer. Finally,  the \textbf{\textit{B2-Pyr}} molecule absorption spectrum represents the envelope of all possible electronic transitions at different absorption wavelengths finding a higher contribution in the electronic transition at $596.508$~nm in Toluene associated with the  \textit{HOMO}$\rightarrow$\textit{LUMO+3} transition, and $590.170$~nm in Methanol corresponding to the \textit{HOMO}$\rightarrow$\textit{L+2}  transition.
\medskip
\noindent
Another parameter that provides information on the radiation capture efficiency of the electron-donor material (\textbf{\textit{BDP-Is}}, \textbf{\textit{BDP-Pyr}}, \textbf{\textit{B2-Is}}, and \textbf{\textit{B2-Pyr}}) in addition to the absorption spectra are the light harvesting efficiencies ($LHE$). Generally, the $LHE$ are closest to the magnitude of the oscillator strength $f_{osc}$, and can be expressed as ~\sugb{\cite{tai2015, manzoor2017}}: 
\begin{equation}
LHE=1-10^{-f_{osc}}.
\label{Ecu4_8C}
\end{equation}
\noindent
According to Eq.~\sugb{\ref{Ecu4_8C}}, the bigger $f_{osc}$, the higher the LHE. The LHE corresponding to $\lambda_{max}$ in the molecules present the following order  $\textbf{\textit{B2-Is}}> \textbf{\textit{B2-Pyr}}> \textbf{\textit{BDP-Is}}> \textbf{\textit{BDP-Pyr}}$, as shown in Table~\sugb{\ref{Tab4_5C}}. The LHE values for the  $D$--$\pi$--$A$ derivatives are in the range between $0.751-0.856$, meaning that all electron-donor compounds have a similar sensitivity to the sunlight~\sugb{\cite{fitri2014,tai2015}}.

\subsection{Electronic Properties} \label{Electronic}

There are several molecular systems that have been used in organic solar cells (OSCs) in order to improve the energy conversion efficiency in those devices. We remark that the theoretical knowledge about the \textit{HOMO} and \textit{LUMO} molecular energies is crucial to understand the electronic dynamics in those cells. The \textit{HOMO} and \textit{LUMO} energies of the donor segments ($D$--$\pi$--$A$) and the \textit{LUMO} levels of acceptor part are relevant parameters to determine the efficient charge transfer between donor and acceptor.  In OSCs, the energy of the \textit{HOMO--LUMO} frontier orbitals of the photo-active components have a close relationship with the photovoltaic properties, because the energy levels are related the to open-circuit voltage ($V_{oc}$) and the energy driving force ($\Delta E$) for the exciton dissociation \sugb{\cite{berube2013}}. For an estimation of the  \textit{HOMO-LUMO} frontier orbitals energy of the molecular complexes we adjusted the equations of B\'erub\'e {\it et al.}~\sugb{\cite{berube2013}}, such that:
\begin{eqnarray}
HOMO_{Est} &=& 0.68\times HOMO_{DFT}-1.92 \; {\textrm{eV}} \\ \nonumber
LUMO_{Est} &=& 0.68\times LUMO_{DFT}-1.59 \;  {\textrm{eV}} 
\end{eqnarray}
\begin{table}[htp] 
\begin{center}
\centering
\caption{Calculated values of the frontier orbitals \textit{HOMO} (\textit{H}) and \textit{LUMO} (\textit{L}) energies which constitute the photo-activated material with the respective estimation (\textit{H}$_{Est}$ and \textit{L}$_{Est}$), the energy gap $\Delta E_{g}$, and the energy of the exciton driving force $\Delta E$ (in eV), for exciton dissociation in Toluene and Methanol.}
\scalebox{0.78}{
\begin{tabular}{ccccccc} 
\hline
\hline
\\  
\centering\textbf{Molecule} & \textbf{HOMO$_{DFT}$} & \textbf{LUMO$_{DFT}$} & \textbf{HOMO$_{Est}$} & \textbf{LUMO$_{Est}$} & \textbf{$\Delta$\textit{E}$_{g}$} & \textbf{$\Delta$\textit{E}} \\
&&&&&&\\
\hline
\hline
\large{\text{Toluene}}  &       &       &        &        &       &       \\
\textbf{\textit{BDP-Is}}  & -5.51 & -3.28 & -5.667 & -3.820 & 1.846 & 0.388 \\
\textbf{\textit{BDP-Pyr}} & -5.47 & -3.14 & -5.640 & -3.725 & 1.914 & 0.483 \\
\textbf{\textit{B2-Is}}   & -5.10 & -3.27 & -5.388 & -3.814 & 1.574 & 0.394 \\ 
\textbf{\textit{B2-Pyr}}  & -5.08 & -3.12 & -5.374 & -3.712 & 1.663 & 0.496 \\  
\textbf{\textit{PCBM}}    & -5.91 & -3.85 & -5.939 & -4.208 & 1.731 & N.A.  \\
                          &       &       &        &        &       &       \\
\hline
\hline
\large{\text{Methanol}} &       &       &        &        &       &       \\
\textbf{\textit{BDP-Is}}  & -5.58 & -3.20 & -5.714 & -3.766 & 1.948 & 0.442 \\
\textbf{\textit{BDP-Pyr}} & -5.55 & -3.10 & -5.694 & -3.698 & 1.996 & 0.510 \\
\textbf{\textit{B2-Is}}   & -5.21 & -3.17 & -5.463 & -3.746 & 1.717 & 0.462 \\ 
\textbf{\textit{B2-Pyr}}  & -5.20 & -3.06 & -5.456 & -3.671 & 1.785 & 0.537 \\
\hline
\hline
\end{tabular}}
\label{Tab_5}
\end{center}
\end{table}
\noindent We performed the calculations by employing the B3LYP/6--311G(d,p) level to estimate the geometrical structure, the frontier orbitals, the exciton driving force energy, and the corresponding band-gap for eight compounds \textbf{\textit{BDP-Is}}, \textbf{\textit{BDP-Pyr}}, \textbf{\textit{B2-Is}}, and \textbf{\textit{B2-Pyr}}, listed in the Table \sugb{\ref{Tab_5}}.
\medskip

The \textit{HOMO}/\textit{LUMO} values of the studied compounds are show in Table~\sugb{\ref{Tab_5}}, and are in good agreement with previously reported experimental work\sugb{\cite{cabrera2017, cabrera2018}}. The [6,6]-phenyl-C$_{60}$--butyric acid methyl ester (abbreviated as PCBM) will be used as the electron-acceptor material, which is an excellent    electron transporter with a \textit{LUMO} energy ($-3.95$ eV~\sugb{\cite{Sharma2012}}) that is both  high  enough to support a large photo-voltage, given that $V_{OC}$ is limited by the energy difference between donor \textit{HOMO} and acceptor \textit{LUMO}, but also low enough to provide ohmic contacts for electron extraction and injection from common cathode electrons~\sugb{\cite{brabec2011}}. The \textit{HOMO} and \textit{LUMO} values of the electron-acceptor component PCBM are experimentally reported in~\sugb{\cite{wu2010,Sharma2012,Huang2013,zhang2014,Jagadamma2015,Yoo2011}}; these are in good agreement with those calculated theoretically in this work and reported in the Table \sugb{\ref{Tab_5}} ($-5.91$ eV/$-3.85$ eV respectively).
\medskip

On the other hand, when  comparing between the  molecules, the calculated band gap $\Delta E_{g} = E_{_{LUMO}} - E_{_{HOMO}}$ increases  in the following hierarchy $\Delta E_{g-\mathbf{\textbf{\textit{BDP-Pyr}}}}>\Delta E_{g-\mathbf{\textbf{\textit{BDP-Is}}}}>\Delta E_{g-\mathbf{\textbf{\textit{B2-Pyr}}}}>\Delta E_{g-\mathbf{\textbf{\textit{B2-Is}}}}$.  Other authors have previously reported the improvement of the photovoltaic properties correlated with a decreasing value  of the energy gap ($\Delta E_{g}$) \sugb{\cite{tai2015}}. Consequently, the \textbf {\textit{B2-Is}} compound is a suitable candidate for a better photovoltaic performance regarding the molecules here analyzed. The low values of $\Delta E_{g-\mathbf{\textbf {\textit{B2-Is}}}}$ compared to $\Delta E_{g-\mathbf{\textbf {\textit{BDP-Is}}}}$, $\Delta E_{g-\mathbf{\textbf {\textit{BDP-Pyr}}}}$, and $\Delta E_{g-\mathbf{\textbf {\textit{B2-Pyr}}}}$ indicates a significant intramolecular charge transfer in \textbf{\textit{B2-Is}}, which translates  into a red-shift of the absorption spectrum.
\medskip

Comparing the band gap between the molecules \textbf{\textit{BDP-Is}} and \textbf{\textit{BDP-Pyr}}, as well as the  \textbf{\textit{B2-Is}} and \textbf{\textit{B2-Pyr}} compounds, we find a larger band gap energy for \textbf{\textit{BDP-Pyr}} and \textbf{\textit{B2-Pyr}} than for \textbf{\textit{BDP-Is}} and \textbf{\textit{B2-Is}}. We can attribute this to the greater conjugation of the ion pair electrons on the oxygen and nitrogen atoms of the isoxazoline fragment with the conjugation $\pi$-system of the BODIPY core. The estimation of the exciton driving force ($\Delta E$) is helpful to predict the degree of efficient charge transfer between the photo-active materials within organic solar cells and is denoted\sugb{\cite{wang2014}} by:
\begin{equation}
\Delta E=LUMO_{Donor}-LUMO_{Acceptor}
\label{Ecu4_9C}
\end{equation}
\noindent
The expression above defines the difference between the \textit{LUMO} orbitals of the electron-donor ($D$--$\pi$--$A$ system) and the electron-acceptor (PCBM) materials. The obtained result shows the energy differences ($\Delta E$) higher than 0.3~eV in all the studied composites. Therefore, an efficient exciton splitting in free charge carriers (electron-hole pairs), as well as electrons transfer between electron-donor and electron-acceptor materials, can be guaranteed. The energy losses in these molecules are minimized due to charge carriers recombination\sugb{\cite{walker2013, wang2014, duan2013}}. In Table~\sugb{\ref{Tab_5}},  a $\Delta E$ range (0.388--0.496~eV) is calculated, and from this result, an ordering between the molecular complexes appear: $\Delta E_{\mathbf{\textbf{\textit{B2-Pyr}}}}>\Delta E_{\mathbf{\textbf{\textit{BDP-Pyr}}}}> \Delta E_{\mathbf{\textbf{\textit{B2-Is}}}}>\Delta E_{\mathbf{\textbf{\textit{BDP-Is}}}}$. The latter can be associated with molecular $\pi$-delocalization link in which the electrons are free to move in more than two nuclei, implying that the $\pi$-delocalization length affects the $\Delta E$ in the $D$--$\pi$--$A$ derivatives ~\sugb{\cite{tai2015}}. In this sense, a longer $\pi$-delocalization length produces a lower $\Delta E$ as in the case for  \textbf{\textit{BDP-Is}} and \textbf{\textit{B2-Is}} molecules (0.388~eV and 0.394~eV, respectively)  in contrast with \textbf{\textit{BDP-Pyr}} and \textbf{\textit{B2-Pyr}} (0.483~eV and 0.496~eV, respectively).  The delocalization is attributable to the oxygen atom inclusion in the donor--$\pi$--acceptor ($D$--$\pi$--$A$) systems, limiting the conjugation and facilitating the electronic $\pi$--delocalization.  Furthermore, we observed that molecules with an oxygen atom in the structure have a better dissociation capability at the electron-donor/electron-acceptor interface due to a lower value of $\Delta E$. The previous results show that the \textbf{\textit{BDP-Is}} and \textbf{\textit{B2-Is}} molecules have a lower $\Delta E$ guaranteeing the effective exciton dissociation which together with the low $E_g$ bring these materials as potential candidates for bulk heterojunction solar cells.

\subsection{Bulk Heterojunction Solar Cell} \label{solar_cell}

\begin{figure*}[ht] 
\centering
\includegraphics[width=\textwidth]{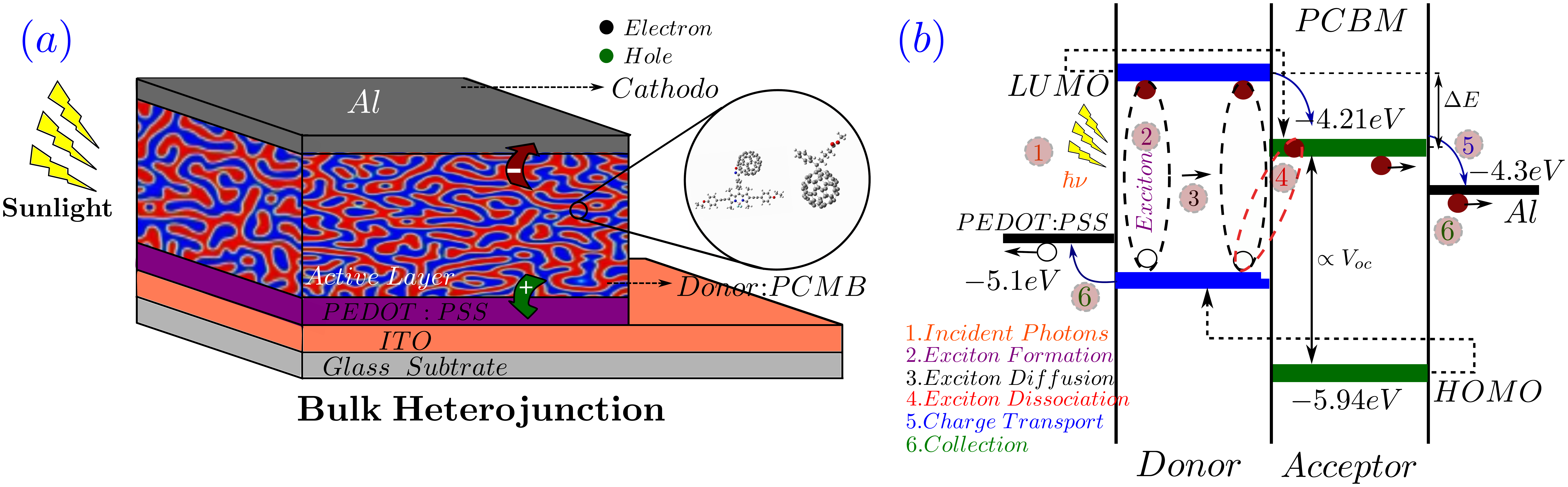}
\caption{The design of a BHJ OSC requires the following, as diagrammatically sketched in the steps indicated for the energy-levels diagram in (a): 1) incident photons, 2) exciton formation, 3) exciton diffusion, 4) exciton dissociation, 5) charge transport, 6) collection.  (b) Configuration of the electron-donor material, the sketched PCBM structure.}
\label{Fig_7}
\end{figure*}

In bulk heterojunction (BHJ) solar cells, the active layer, which consists of two kinds of molecular materials (the electron-donor and the electron-acceptor material), is sandwiched between two electrodes with different work functions, as shown Figure~\sugb{\ref{Fig_7}}.
\medskip
To understand the operational principle of a solar cell, we divide it into six different sub-processes, as explained through the energy band-gap diagram of the donor and acceptor materials of  Figure~\sugb{\ref{Fig_7}}. The steps are as  follows: (1) the active layer absorbs photons producing electronic transitions between the \textit{HOMOs} and \textit{LUMOs} states of the donor material, (2) the electronic transition generates a system of electron-hole pairs known as an excitons, (3) the excitons are diffused toward the acceptor material interface producing excitonic dissociation forming a charge transfer (CT) complex, which will be favorable to occur when the energy difference between the donor \textit{LUMO} and the acceptor \textit{LUMO} ($\Delta E$) is greater than the binding energy of the exciton~\sugb{\cite{wang2014}}, (4) if the distance between the electron and hole becomes greater than the Coulomb trapping radius, the charge transfer state becomes a separated (CS) state (or free charge carriers) in the photovoltaic process. However, if the electrons are unable to escape from the coulomb trapping radius, the geminate pair will recombine across the donor/acceptor interface constituting another losing mechanism in these devices~\sugb{\cite{mazzio2014}}, (5) the dissociated charges are transported through $p$-type or $n$-type domains toward the metallic electrodes. Finally, (6) the collected holes at the anode and electrons at the cathode can be employed in external circuits~\sugb{\cite{Gao2014}}. While the exciton dissociation process is actually far more complex than depicted, these simplified schematics are useful for generating a conceptual understanding of the photophysical processes occurring in the OPVs.

\subsection{Photovoltaic Properties} \label{Photovoltaic}

Aiming to provide a working guide that can be useful for experimental researchers in the field, we used the Scharber's model to estimate the photovoltaic properties of the analyzed materials. Initially, the  bulk heterojunction organic solar cells (BHJ OSC) present a mixture of an electron-donor $\pi$-conjugate with an electron-acceptor derivative from fullerene. Therefore, we examine the photovoltaic behavior of the  new \textit{D--$\pi$--A} derivatives   mixed whith PCBM (see Fig.~\sugb{\ref{Fig_1}}), as a widely used electron-acceptor  in solar cell devices~\sugb{\cite{dennler2009, zheng2016, mangaud2017, lee2017, wang2018}}. We evaluated the power conversion efficiency ($PCE$) as the most commonly used parameter to compare the performance of many solar cells, as well as some important parameters such as the short-circuit current density ($J_{sc}$), the open circuit voltage ($V_{oc}$), and the fill factor ($FF$). The photoelectric conversion efficiency of solar cell devices under sunlight irradiation (e.g. AM1.5G) can be determined by  $J_{sc}$ and  $V_{oc}$, and the power of incident light $P_{in}$, as follows~\sugb{\cite{lu2015}}:
\begin{equation}
\eta=\frac{P_{out}}{P_{in}}=\frac{V_{oc}J_{sc}FF}{P_{in}}\times100,
\label{Ecu4_10C}
\end{equation}
\noindent where the fill factor ($FF$) is proportional to the maximum power of the solar cell. This approach allows us to estimate the $PCE$ value of a given polymer or molecular systems in the active layer from its frontier orbital energy levels. The frontier energy  levels and their separation ($E_g$) are critical for  photovoltaic behavior,  since they directly affect the $J_{sc}$ and $V_{oc}$ values, then a proper control of those levels is of prime importance in the design of high-performance photovoltaic devices~\sugb{\cite{facchetti2010}}. 
\begin{table}[htp] 
\begin{center}
\centering
\caption{Parameter values of the studied molecular compounds: Energy gap ($E_{g}$), open circuit voltage ($V_{oc}$), short-circuit current density ($J{sc}$), fill factor ($FF$), and power conversion efficiency ($PCE$).}
\scalebox{0.78}{
\begin{threeparttable}
\begin{tabular}{ccccccc} 
\hline \hline
\\
 Molecule & $\Delta E_{g}$ (eV) & $V_{oc}$ (V) & $J_{sc}$ (mA/cm$^{2}$) & $FF$ & PCE (\%) & PCE (\%)\tnote{a}
\\ 
\hline \hline
\text{Toluene}          &       &       &        &       &        &        \\
\textbf{\textbf{BDP-Is}}  & 1.846 & 1.159 & 11.939 & 0.895 & 12.380 & 9.523 \\ 
\textbf{\textbf{BDP-Pyr}} & 1.928 & 1.132 & 10.575 & 0.893 & 10.685 & 8.219 \\  
\textbf{\textbf{B2-Is}}   & 1.574 & 0.880 & 17.155 & 0.870 & 13.136 & 10.105 \\
\textbf{\textbf{B2-Pyr}}  & 1.663 & 0.866 & 15.353 & 0.869 & 11.555 & 8.888 \\
                          &       &       &        &       &        &        \\   \hline \hline
\text{Methanol}         &       &       &        &       &        &        \\
\textbf{\textbf{BDP-Is}}  & 1.948 & 1.206 & 10.248 & 0.898 & 11.103 & 8.541 \\
\textbf{\textbf{BDP-Pyr}} & 1.996 & 1.186 & 9.541  & 0.897 & 10.146 & 7.805 \\
\textbf{\textbf{B2-Is}}   & 1.717 & 0.955 & 14.295 & 0.878 & 11.984 & 9.218 \\
\textbf{\textbf{B2-Pyr}}  & 1.785 & 0.948 & 13.024 & 0.877 & 10.832 & 8.332 \\
\hline
\end{tabular}
\begin{tablenotes}
\item[a] The PCE value was calculated by considering an EQE of 50 \%.
\end{tablenotes}
\end{threeparttable}}
\label{Tab_6}
\end{center}
\end{table}
\medskip

The maximum open circuit voltage ($V_{oc}$) of the BHJ solar cell according to Scharber's model is related to the difference between the highest occupied molecular orbital (\textit{HOMO}) in the electron-donor and the \textit{LUMO} of the electron-acceptor materials regarding the energy lost during the photo-charge generation~\sugb{\cite{scharber2006}}. The theoretical value of open-circuit voltage $V_{oc}$ of a conjugated polymer-PCBM solar cell can be estimated by~\sugb{\cite{scharber2006}}:
\begin{equation}
eV_{oc}=\vert E_{HOMO}^{Donor}\vert-\vert E_{LUMO}^{PCBM}\vert-0.3~\text{eV},
\label{Ecu4_11C}
\end{equation}
\noindent where $e$ is the elementary charge and 0.3~eV is an empirical factor. In operational solar cells, additional recombination paths decrease the value of $V_{oc}$ affecting the PCE of  OPVs because they must operate at a voltage lower than the $V_{oc}$. A decreasing $V_{oc}$  implies a power reduction ($J_{sc}\times V_{oc}$) given by the fill factor ($FF$).  A simple relation between the energy level of the \textit{HOMO} of the molecular systems (\textbf{\textit{BDP-Is}}, \textbf{\textit{BDP-Pyr}}, \textbf{\textit{B2-Is}}, and \textbf{\textit{B2-Pyr}}) and the $V_{oc}$ is derived to estimate the maximum efficiency of the BHJ OSCs.  Based on these considerations, the ideal parameters for a $\pi$-conjugated-system-PCBM device are determined in Table~\sugb{\ref{Tab_6}}. The calculated $V_{oc}$ is in a range between 0.886~V to 1.159~V in Toluene and from 0.948~V to 1.206~V in Methanol, these are relatively high values implying that the OPV composed by the studied molecule as electron-donor material and PCBM has potential application as BHJ OSCs because of their  improved $V_{oc}$ . The  $V_{oc}$ hierarchy goes as follows: $V_{oc}^\mathbf{\textbf{\textit{BDP-Is}}} >V_{oc}^\mathbf{\textbf{\textit{BDP-Pyr}}} >V_{oc}^\mathbf{\textbf{\textit{B2-Is}}} >V_{oc}^\mathbf{\textbf{\textit{B2-Pyr}}}$  in both of the solvents, showing for the  \textbf{\textit{BDP-Is}} and \textbf{\textit{BDP-Pyr}} molecules values higher than 1.0~V, which means an efficient increasing of the $V_{oc}$.
\medskip

Given that the $PCEs$ are estimated from the values for the open circuit voltage $V_{oc}$, the short circuit current $J_{sc}$, and the fill factor $FF$ of the OSCs~\sugb{\cite{guo2012}}, to achieve high efficiencies, an ideal donor material should have a low energy gap and a deep \textit{HOMO} energy level (thus increasing $V_{oc}$). Additionally, a high hole mobility is also crucial for the carrier transport to improve $J_{sc}$ and $FF$, which can be empirically described  by~\sugb{\cite{nayak2012, green1982}}
\medskip
\begin{equation}
FF=\frac{\nu_{m}}{\nu_{m}+1}\frac{\nu_{oc}-\ln(\nu_{m}+1)}{\nu_{oc}(1-\e^{-\nu_{oc}})},
\label{Ecu4_12C}
\end{equation}
\noindent with $\nu_{oc}=\frac{V_{oc}}{n\kappa_{B}T}$, and $\nu_{m}=\nu_{oc}-\ln(\nu_{oc}+1-\ln(\nu_{oc}))$. The parameter values corresponding to the \textbf{\textit{BDP-Is}}, \textbf{\textit{BDP-Pyr}}, \textbf{\textit{B2-Is}}, and \textbf{\textit{B2-Pyr}} compounds are shown in Table~\sugb{\ref{Tab_6}}. The fill factor oscillates between 0.869 to 0.895 in Toluene, and from $0.877$ to $0.898$ in Methanol, in agreement with the reported BHJ SC values for different polymers as electron-donor, and PCBM as electron-acceptor material~\sugb{\cite{huo2015,trukhanov2015}}.
\medskip

If the energy conversion efficiency is calculated with Eq.~\sugb{\ref{Ecu4_10C}}, by taking the $V_{oc}$ and $FF$ from the Scharber's model, and considering  experimental $J_{sc}$ values, a correlation between the calculated $PCE$ and the experimental $PCE$ value showing an absolute standard deviation of 0.8\% is found~\sugb{\cite{hou2008}}. However, if the theoretical value for $J_{sc}$ and experimental $FF$ is used to calculate the PCE, a considerable change is observed in the correlation between $PCE_{Ter}$ and $PCE_{Exp}$, which indicate a high sensibility of $J_{sc}$ to the structural or morphology change in the OPV devices. Thus, similar to the $FF$ and $V_{oc}$, the short-circuit current density ($J_{sc}$) is another important parameter involved in the calculation of $PCE$ and can be determined as follows~\sugb{\cite{dennler2009}}:
\begin{equation}
J_{sc}=q\int_{\lambda_{1}}^{\lambda_{2}}\phi_{AM1.5G}(\lambda)\times EQE(\lambda) d\lambda,
\label{Ecu4_13C}
\end{equation}
\noindent where $EQE(\lambda)$ is the external quantum efficiency, $\phi_{AM1.5G}$ is the flow of photons associated with the solar spectral irradiance in AM1.5G, $q$ is the electronic charge, and $\lambda_{1}$ and $\lambda_{2}$ are the limits of the active spectrum of the device. The photon absorption rate of the donor and acceptor materials, the exciton dissociation efficiency, and the resultant charge transportation efficiencies toward the electrodes limit the $J_{sc}$ values. This quantity represents the probability for a photon of being converted into an electron. 
\medskip

The highest reported $EQEs$ are around  50$\%$--65$\%$~\sugb{\cite{Jagadamma2015, Scharber2013a}}. The calculated values reported in Table~\sugb{\ref{Tab_6}}  for $PCE$ and $J_{sc}$ have considered an  $EQE$ of 65\%. These values show that the \textbf{\textit{BDP-Is}} and \textbf{\textit{B2-Is}} compounds have higher PCE probably associated with the insertions of an oxygen atom in their molecular structure. Additionally, we can observe that the \textbf{\textit{B2-Is}} molecule is a great candidate as a compound in the manufacture of organic solar cells. 

We can further observe that although the  \textbf{\textit{BDP-Is}} and \textbf{\textit{BDP-Pyr}} systems possess a comparable $V_{oc}$ of $1.159$~V and $1.132$~V in Toluene, respectively, and a quite similar form factor, the $J_{sc}$ is completely different, marking a great difference in their corresponding $PCE$ values
\medskip

Since the model uses rather simple assumptions for the $EQE$ value in Eq.~\sugb{\ref{Ecu4_13C}}, there is an associated imprecision in the $J_{sc}$ estimations. The $EQE$ is a complex frequency-dependent function that considers the transport effects and the morphology of the films.  The $J_{sc}$ values from the Scharber's model do not define a strict upper bound and should be considered only as an indicator for the molecular capability as a light harvesting device. In this way, the estimation of the  $EQE$  value is a sensitive parameter in Sharber's model to define the accuracy between the calculated against the actual experimental PCE values.

\subsection{Molecular systems as electron-acceptor materials} \label{Acceptor_material}

The $C_{60}$ fullerene has often been chosen as an excellent electron-acceptor since it presents a  triple-degenerated low-energy \textit{LUMO}.  This molecular structure is also capable of reversibly accepting up to six electrons and offers unusually low reorganization energy in charge transfer processes allowing ultra-fast charge separation processes and slow charge recombination~\sugb{\cite{guldi2000, konev2015}}. Therefore, we expect that the merger between the $C_{60}$ and the BODIPY unit may exhibit acceptor properties being able to act as electron-acceptor materials in molecular photovoltaic devices. Therefore, here we consider the analyzed systems as electron-acceptors materials, due to the presence of the $C_{60}$ fullerene in their molecular structure. Hence, we consider the Poly(3--hexylthiophene--2,5--diyl)~(\textbf{\textit{P3HT}}) as electron-donor material because of the relative stability, ease of scalability by direct synthesis~\sugb{\cite{ludwigs2014}}, and compatibility with high-performance production techniques. Additionally, the widespread use and the well known  material capabilities in OPV research during some time, make the \textbf{\textit{P3HT}} a typical candidate to perform some theoretical tests~\sugb{\cite{mulligan2014, po2014}}. The \textbf{\textit{P3HT}} became in the pioneering material to research on conjugated polymers due to advances in synthetic methodologies,  establishing applications in several organic electronic devices such as solar cells, field effect transistors, light emitting diodes, and many others. This material has commonly been employed in several fundamental studies regarding charge transport and film morphology because of the smoothness of the synthesis and the high-grade opto-electronic properties~\sugb{\cite{marrocchi2012, sista2014}}. \textbf{\textit{P3HT}} is soluble in a variety of solvents allowing advantages over other electron-donor materials as the reduced band gap and the high mobility of holes  ($>0.1$~cm$^2$/Vs) with a suitable morphology control and an absorption edge in $650$~nm which matches with the maximum solar photon flux between $600$~nm and $700$~nm~\sugb{\cite{berger2018}}. In this work, the molecular orbitals \textit{HOMO}/\textit{LUMO} ($5.10$~eV/$2.65$~eV) values were calculated following the section ``computational details'', and are in good  agreement with the reported values in the literature~\sugb{\cite{tremel2014, manna2015, gmucova2015}}.

\begin{table}[htp] 
\begin{center}
\centering
\caption{Photovoltaic properties  of the molecular compounds: Energy of the exciton driving force $\Delta E$, open circuit voltage $V_{oc}$, short-circuit current density $J_{sc}$, fill factor $FF$, and  power conversion efficiency PCE.}
\scalebox{0.85}{
\begin{threeparttable}
\begin{tabular}{ccccccc} 
\hline \hline
\\ 
 Molecule & $\Delta E$ & $V_{oc}$ (V) & $J_{sc}$ (mA/cm$^{2}$) &  $FF$  &  PCE (\%) & PCE (\%)\tnote{a}
\\ 
\hline   \hline
\textbf{\textit{BDP-Is}}  & 0.428 & 1.268 & 9.541 & 0.902 & 10.527 & 8.098 \\ 
\textbf{\textit{BDP-Pyr}} & 0.320 & 1.376 & 9.541 & 0.908 & 11.543 & 8.879 \\  
\textbf{\textit{B2-Is}}   & 0.422 & 1.274 & 9.541 & 0.902 & 10.590 & 8.146 \\
\textbf{\textit{B2-Pyr}}  & 0.354 & 1.342 & 9.541 & 0.906 & 11.225 & 8.635 \\ 
\hline
\end{tabular}
\begin{tablenotes}
\item[a] The PCE value was calculated by considering an EQE of 50 \%.
\end{tablenotes}
\end{threeparttable}}
\label{Tab_7}
\end{center}
\end{table}
\noindent Table~\sugb{\ref{Tab_7}} shows the calculated photovoltaic properties values considering the Scharber's method~\sugb{\cite{scharber2006}} in which the active layer is a P3HT: 
$D$--$\pi$--$A$ system. We observed that all calculated $V_{oc}$ show large values (1.268~V--1.376~V) implying that the $D$--$\pi$--$A$ systems improved the P3HT $V_{oc}$, thus opening the possibility for photovoltaic applications. In addition, we observed a range around 0.320~eV--0.428~eV in the difference of the \textit{LUMO} energy levels between \textbf{\textit{P3HT}} and $D$--$\pi$--$A$ systems,  suggesting a photo-excited electron transfer from the \textbf{\textit{P3HT}} to $D$--$\pi$--$A$ fair enough to be employed in photovoltaic devices. In this order of ideas, we observed lower  $\Delta E$ values for the \textbf{\textit{BDP-Pyr}} and \textbf{\textit{B2-Pyr}} systems involving most probable dissociation processes in the donor/acceptor interface than in \textbf{\textit{BDP-Is}} and \textbf{\textit{B2-Is}} molecules.

Therefore, a lower $\Delta E$ not only ensures the dissociation of the effective exciton but also reduces the energy loss by recombination processes. Thus, molecular systems (electron-acceptor molecules) with high $V_ {oc}$ and low $\Delta E$ should improve the solar cell performance.  The low-band-gap of the alkoxyphenylethynyl group allows to increase the PCE  by the light absorption up to the infrared range (to harvest additional solar energy), and also changes the \textit{HOMO} and \textit{LUMO}  levels of the analyzed molecules~\sugb{\cite{kroon2008}}. Based on Scharber's model~\sugb{\cite{scharber2006}}, the maximum PCE of the photovoltaic solar cells with P3HT: $D$--$\pi$--$A$ as active layer is around 11.54\% for the \textbf{\textit{BDP-Pyr}} system and 11.23\% for the \textbf{\textit{B2-Pyr}} molecule as electron-acceptor materials. 
\medskip

Although the results chosen by the PCE with the method presented here do not represent the value they would achieve in real experimental  conditions, but instead the maximum value reached, some works~\sugb{\cite{kar2017,lopez2017,berube2013}} show that the decrease in the PCE factor that can be achieved when comparing the results both theoretical and experimental is around 30\%--50\%. Which shows that our systems are robust and could be of interest in the construction of OSCs or opto-electronic devices according to the PCE results shown by the NREL cell efficiency chart~\sugb{\cite{NREL2019}}.

\section{Conclusion} \label{Conclusion}

The energy estimation of the frontier molecular orbitals \textit{HOMO}--\textit{LUMO} showed a high correlation with the available experimental data.   The energy of the exciton driving force ($\Delta E$) of all the  $D$--$\pi$--$A$ derivatives has a value greater than 0.3 eV, which guarantees an efficient  exciton dissociation. The results here presented demonstrate the crucial role played by the donor \textit{LUMO} level in bulk-heterojunction solar cells: besides a reduction of the band-gap, new donor materials should be designed to optimize their \textit{LUMO} value because this parameter drives the solar-cell efficiency.  As seen in Table~\sugb{\ref{Tab_6}}, we remark that an optimized open-circuit voltage translates into optimized device conversion efficiencies. Comparing the \textbf{\textit{BDP-Is}} and \textbf{\textit{BDP-Pyr}} systems, as well as the  \textbf{\textit{B2-Is}} and \textbf{\textit{B2-Pyr}} compounds, we found a favorable dissociation of the exciton in the donor/acceptor interface by the  electrons conjugation of the ion pair in the oxygen and nitrogen atoms (of the isoxazoline fragment) with the $\pi$-conjugated BODIPY core system; consequently, a favorable PCE in the \textbf{\textit{DBP-Is}}:\textbf{\textit{PBCM}}, and \textbf{\textit{B2-Is}}:\textbf{\textit{PBCM}} devices.  Our results, for the proposed molecular compounds, show that the system with  lowest  $\Delta E$, and  highest  $V_{oc}$ values exhibit excellent photo-activation features, and therefore are very well suited candidates for BHJ solar cell implementation.
\medskip

Finally, when molecular systems are considered as electron-acceptor materials, due to the presence of fullerene in their structure, it is found that \textbf{\textit{BDP-Pyr}} and \textbf{\textit{B2-Pyr}} have a favorable excitons dissociation and transport of holes than the other systems, hence they would be the most promising ones for application as electron-acceptor materials in organic solar cells, which is also reflected in their PCE values. In addition, we found that the inclusion of a pyrrolidine ring favors the PCE by comparing the \textbf{\textit{BDP-Is}} and \textbf{\textit{B2-Is}} systems.  However, the addition of the alkoxyphenylethynyl group does not represent a favorable change for neither systems. We finally remark that the  studied molecular systems exhibit properties that are favorable for application as photovoltaics if these are used in conjunction with \textbf{\textit{PCBM}} and \textbf{\textit{P3HT}}, according to the needs, as electron-donor or electron-acceptor materials.
\medskip

In summary, this work makes a theoretical contribution to the physical-chemistry of photovoltaic materials by establishing criteria for power conversion efficiency optimization in D--$\pi$--A molecular complexes and BHJ organic solar cells. The performed calculations range from the single molecule domain to the PV properties of BHJ OSCs with novel electron-donor systems, which are a rapidly evolving photovoltaic technology with a projected significant role in the PV-market.  This said, substantial research and development efforts are still needed to achieve the performance level required to become a commercially competitive alternative.

\section*{Acknowledgements}

D. M.-Ú. thanks A.-G. Mora-León for discussions about structure and chemical processes. This work was supported by the Colombian Science, Technology and Innovation Fund-General Royalties System---``Fondo CTeI-Sistema General de Regal\'ias'' under contract No. BPIN 2013000100007, and Vicerrectoría de Investigaciones at Universidad del Valle (Grants CI 9517-2 and  CI 71212).

\section*{Supporting Information}

The Supporting Information is available  available free of charge on the ACS Publications website.

\begin{itemize}
\item Four molecular compounds, with electronic, optical, and photovoltaic properties, and a detailed description about fifteen features.
\end{itemize}

\bibliographystyle{ieeetr}
\bibliography{Bibliography}

\end{document}